\documentclass[
 preprint,
 amsmath,amssymb,
 12pt,
 final,
 tightenlines,
 nofloats,
 floatfix,
 nofootinbib,
 superscriptaddress,
 showkeys,
 showkeywords,
 ]
{revtex4-2}

\usepackage[T2A]{fontenc}
\usepackage[utf8x]{inputenc}
\usepackage[russian,english]{babel}
\usepackage{graphicx}
\usepackage{dcolumn}
\usepackage{bm}
\usepackage{amsmath}
\usepackage{color}

\input{maik.rty}

\setcitestyle{authoryear,round}
\def\squareforqed{\hbox{\rlap{$\sqcap$}$\sqcup$}}

\def\sq{\ifmmode\squareforqed\else{\unskip\nobreak\hfil
\penalty50\hskip1em\null\nobreak\hfil\squareforqed
\parfillskip=0pt\finalhyphendemerits=0\endgraf}\fi}

\def\utw{\smash{\rlap{\lower5pt\hbox{$\sim$}}}}

\def\udtw{\smash{\rlap{\lower6pt\hbox{$\approx$}}}}

\def\diameter{{\ifmmode\mathchoice
{\ooalign{\hfil\hbox{$\displaystyle/$}\hfil\crcr
{\hbox{$\displaystyle\mathchar"20D$}}}}
{\ooalign{\hfil\hbox{$\textstyle/$}\hfil\crcr
{\hbox{$\textstyle\mathchar"20D$}}}}
{\ooalign{\hfil\hbox{$\scriptstyle/$}\hfil\crcr
{\hbox{$\scriptstyle\mathchar"20D$}}}}
{\ooalign{\hfil\hbox{$\scriptscriptstyle/$}\hfil\crcr
{\hbox{$\scriptscriptstyle\mathchar"20D$}}}}
\else{\ooalign{\hfil/\hfil\crcr\mathhexbox20D}}%
\fi}}

\newcommand{\aap}{Astron. and Astrophys. }

\renewcommand{\apj}{Astrophys.~J. }
\newcommand{\apjs}{Astrophys.~J. Suppl. }
\newcommand{\apss}{Astrophys. and Space Sci. }
\newcommand{\araa}{Annual Rev. Astron. Astrophys. }

\newcommand{\mnras}{Monthly Notices Royal Astron. Soc. }

\newcommand{\pasj}{Publ. Astron. Soc. Japan }

\newcommand{\apjl}{Astrophys.~J.}
\newcommand{\aapr}{Astron. and Astrophys.}

\newcommand{\solphys}{Sol. Phys. }

\begin{document}
\selectlanguage{english}
\title{ELECTRON ACCELERATION AND PLASMA HEATING IN AN IMPULSIVE CONFINED C-CLASS SOLAR FLARE}

\author{\firstname{I.~N.}~\surname{Sharykin}}
 \email{ivan.sharykin@phystech.edu}
\author{\firstname{I.~V.}~\surname{Zimovets}}
\author{\firstname{G.~V.}~\surname{Koynash}}

 \affiliation{Space Research Institute of the Russian Academy of Sciences (IKI RAS), Moscow, Russia}
\author{\firstname{E.~F.}~\surname{Ivanov}}
\author{\firstname{V.~V.}~\surname{Fedenev}}
\author{\firstname{S.~A.}~\surname{Anfinogentov}}

\affiliation{Institute of Solar-Terrestrial Physics, Siberian Branch of the Russian Academy of Sciences (ISTP SB RAS), Irkutsk, Russia}
 
\begin{abstract}
A detailed multi-wavelength analysis of the impulsive C2.8 solar flare SOL2023-03-19T02:12 is presented, focusing on the microwave\,(MW) and X-ray domains. The flare was selected because of its impulsive nature, the relatively simple magnetic morphology of its parent active region\,(AR) NOAA\,13256\,(Eho,~$\beta$), its confined (non-eruptive) evolution, characterised by the absence of both a coronal mass ejection and significant large-scale plasma motions, its moderate intensity, pronounced non-stationary temporal behaviour, and the availability of a unique multi-wavelength dataset. This dataset includes MW spectral observations in the frequency range 2.8--12\,GHz obtained with the new Solar Radio Spectropolarimeter (SOLARSPEL), together with MW imaging observations from the Siberian Radioheliograph (SRH). The flare was also observed by two imaging X-ray telescopes, Solar Orbiter/STIX and ASO-S/HXI. By combining observations across multiple wavelength ranges, we investigate the temporal evolution of plasma heating, electron acceleration, quasi-periodic pulsations\,(QPPs), and the evolving morphology of emission sources throughout different layers of the solar atmosphere. Nonlinear force-free field\,(NLFFF) extrapolations are used to reconstruct the three-dimensional\,(3D) magnetic configuration of the AR, allowing the observed emission patterns to be related to the underlying magnetic topology. We present evidence for a direct coupling between the thermal plasma and the non-thermal electron population in the frame of collisionless plasma during the initial flare stage. The empirical relationships established between the parameters of the accelerated-electron spectrum and those of the thermal plasma during the rise phase of the impulsive stage allow us to estimate the efficiency of the acceleration process (i.e. the probability of a thermal electron being accelerated) under conditions where flare hydrodynamic processes are still poorly developed. We further discuss the influence of chromospheric evaporation on electron acceleration during the initial impulsive phase and suggest that the acceleration efficiency may be directly modulated by the inflow of evaporated chromospheric plasma. Analysis of the extrapolated magnetic field indicates that the flare onset was associated with a system of low-lying sheared magnetic loops located along the polarity inversion line\,(PIL) near a region where the horizontal gradient of the radial magnetic-field strength reached\,$|\nabla_h B_r|\approx 1\,\mathrm{kG\cdot Mm^{-1}}$. Given the confined nature of the event and the reconstructed magnetic configuration, we infer that magnetic reconnection most likely occurred within current sheets possessing a substantial guide-field component. The observed non-stationary QPPs in the non-thermal emission, with periods decreasing from approximately 15 to 9\,s, are interpreted as signatures of a sequence of magnetic reconnection episodes occurring in different magnetic structures and triggered quasi-periodically by a process that remains uncertain, but which may involve propagating slow magnetoacoustic waves.

\end{abstract}
\maketitle

\section{INTRODUCTION}
C-class solar flares, despite their relatively low energy output, play an important role in studies of dynamic processes in the solar atmosphere. Long-term observational statistics show that C-class events occur significantly more frequently than M- and X-class flares and are detected more often than weaker B- and A-class events, many of which remain below the sensitivity threshold of current observational instruments~\cite[]{Xiong2021ASS, Biasiotti2025SoPh}. C-class flares are accompanied by enhanced electromagnetic emission over a broad spectral range and exhibit a wide variety of magnetic configurations and energy-release scenarios within AR, extending beyond the framework of the standard two-ribbon flare model (CSHKP)~\cite[]{PriestForbes2002, Benz2017}. In particular, flare energy release may involve mechanisms such as the coalescence of current-carrying magnetic loops~\cite[]{Tajima1987, Sakai_deJager1996}, current-driven loop structures without magnetic reconnection~\cite[]{Zaitsev2008ENG}, or reconnection processes occurring within current sheets without a significant contribution from large-scale eruptive activity~\cite[]{Podgorny2007, Pariat2009}. Therefore, C-class flares should not be regarded merely as scaled-down versions of more energetic M- and X-class events.

The study of C-class solar flares provides an opportunity to investigate the fine details of magnetic reconnection, particle acceleration, and plasma heating. On the one hand, the comparatively low energy release of such events, relative to M- and X-class flares, results in weaker electromagnetic emission and reduces the likelihood of detector saturation, thereby facilitating studies of the spatial structure and dynamics of flare sources. On the other hand, C-class flares often exhibit complex multi-peaked temporal profiles with QPPs, similar to those observed in more energetic events~\cite[]{Pugh2017, Hayes2020, Szaforz2025}. They may also display other characteristic signatures of flare energy release, such as reversals of the circular-polarization sign in the MW range~\cite[]{AltyntsevMeshalkina2017, BogodYasnov2009}.

Investigations of C-class solar flares provide valuable insights into the interaction of magnetic loops and the mechanisms responsible for the redistribution of released magnetic energy between accelerated particles, heated plasma, and various magnetohydrodynamic wave modes. Such studies are important for advancing our understanding of solar activity and improving space-weather prediction capabilities, since even relatively weak flares may produce measurable effects in the Earth's upper atmosphere and impact satellite-based technologies~\cite[]{Huang2025Atmos, Miteva2023Astro}.

The aim of this study is to perform a comprehensive multi-wavelength investigation of the non-stationary energy release during the C2.8-class solar flare that occurred on 19 March 2023 (02:12/02:15/02:19 UTC) in the penumbral region of the leading sunspot of AR NOAA 13256 (S23E58). The flare was observed by numerous ground- and space-based instruments, including the newly developed Solar Radio Spectropolarimeter (SOLARSPEL) and the Siberian Radioheliograph (SRH), which together provided spectral and imaging MW observations over the frequency range 3--12\,GHz. The event was selected because it is among the first solar flares observed with SOLARSPEL and provides an opportunity both to demonstrate the capabilities of this new instrument and to investigate several important aspects of flare energy release:
\begin{itemize}
\item The MW burst lasted less than one minute, making the event highly impulsive. For such events, the non-stationary background can be determined more reliably than for long-duration flares. This minimizes uncertainties associated with the subtraction of the pre-flare emission and allows a more accurate reconstruction of the intrinsic MW spectrum of the flare source for subsequent modelling.
\item The flare is exceptionally well covered by multi-wavelength observations from numerous instruments operating across different spectral domains. A particularly important advantage is the availability of spatially resolved MW observations from the Siberian Radioheliograph (SRH) together with simultaneous X-ray imaging observations from two independent telescopes, STIX onboard Solar Orbiter and HXI onboard ASO-S. This unique combination enables a detailed investigation of the spatial morphology of the emission sources.
\item The moderate intensity of the flare provides favourable observing conditions because detector saturation is avoided. This is particularly important for modern radio-interferometric imaging techniques, ultraviolet~(UV) image analysis, and X-ray spectral diagnostics.
\item Although C-class flares are often regarded as geometrically simple events, the present flare exhibits a complex multi-peaked temporal profile with QPPs and pronounced non-stationary emission dynamics characteristic of substantially more energetic flares.
\item The flare occurred in a morphologically simple AR, classified as \textit{Eho} (\textit{Hsx} one day earlier) according to the McIntosh classification and as $\beta$ ($\alpha$ one day earlier) according to the Hale magnetic classification. The AR was dominated by a single large sunspot surrounded by only a few small pores. Despite the complex temporal evolution of the flare, the relatively simple magnetic topology of the parent AR facilitates the interpretation of the spatial structure of the non-stationary energy release.
\item The flare was confined and showed no evidence of an eruption. This allows the analysis to neglect the energy associated with large-scale plasma motions and to focus primarily on the two dominant energy channels: particle acceleration and plasma heating. Consequently, the relationship between the thermal and non-thermal components of the flare energy release can be investigated in greater detail. The absence of type II and type III radio bursts in the metric and decimetric ranges, according to observations from the e-Callisto network, independently supports the predominantly closed magnetic configuration of the event.
\end{itemize}

With an estimated total energy release of $\sim10^{27}$--$10^{30}$\,erg, the present event falls into the microflare category~\cite[]{Hannah2011}. Detailed studies of microflares exhibiting QPPs, particularly those accompanied by spectropolarimetric MW observations and/or spatially resolved MW imaging, remain relatively scarce. For example, \cite{Nakariakov2018_micro} reported QPPs with periods of approximately $\approx$~1.4 and $\approx$~0.7\,s in a B2-class microflare and suggested that the observations could be explained by sausage-mode oscillations of a coronal loop, while also noting that alternative mechanisms, including oscillatory coalescence of current-carrying loops or magnetoacoustic oscillations near a magnetic null point, could not be excluded. \cite{Altyntsev2022_double} analysed a C8.3-class flare exhibiting pronounced QPPs with periods of approximately $\approx$~8 and $\approx$~3\,s and interpreted them in terms of energy-release modulation in an oscillating current sheet during loop coalescence. \cite{Kashapova2021} detected QPPs with periods of about $\sim$~30\,s in MW emission and about 6\,s in decimetric emission from a C6.9-class flare, interpreting them as signatures of sausage-mode oscillations together with third-harmonic modulation of magnetic reconnection during the interaction of the flaring loop with surrounding larger loops. QPPs with periods of about $\sim$~1\,s observed simultaneously in X-ray and MW emission from a C8.2-class flare were interpreted by \cite{LiDong2025MNRAS} as signatures of intermittent magnetic reconnection modulated by instabilities arising from interactions between current-carrying loops and magnetic islands (plasmoids). These studies demonstrate that there is still no consensus regarding the physical mechanism responsible for QPPs in weak solar flares, consistent with the situation for larger eruptive events~\cite[see the reviews by][]{Kupriyanova2020, Zimovets2021, Reale2026}. The present multi-wavelength analysis of the C2.8-class flare expands the currently available sample of weak flares exhibiting QPPs and further illustrates the challenges involved in identifying the physical origin of these periodic pulsations.

The relatively compact loop system (approximately $\approx$~10\,Mm) in the selected event provides an opportunity to investigate the relationship between plasma heating and electron acceleration under conditions where the magnetic geometry is not complicated by the large-scale flare arcades typically associated with M- and X-class events. In particular, this study focuses on the following aspects of flare energy release:

\begin{itemize}
\item the properties and physical nature of non-stationary energy release in an impulsive C-class flare occurring in a morphologically simple AR;
\item the morphology of the MW and X-ray emission sources, their relationship to the reconstructed magnetic topology, and the implications for the spatial organisation of high-energy processes;
\item the temporal evolution of the thermal plasma and accelerated electron populations, including the analysis of QPPs, the energy partition between the thermal and non-thermal components, and the properties of the particle acceleration process.
\end{itemize}

The remainder of this paper is organised as follows. Chapter\,2 describes the observational data, instrumentation, and analysis methods. Chapter\,3 presents the multi-wavelength observations and their temporal and spectral characteristics. Chapter\,4 analyses the magnetic-field configuration. Chapter\,5 presents the MW and X-ray spectral analysis. Chapter\,6 discusses the physical interpretation of the results, including the origin of the observed QPPs and the remaining observational and theoretical challenges. Finally, chapter\,7 summarises the main conclusions of this work.

\section{OBSERVATIONS, INSTRUMENTATION, AND DATA ANALYSIS}

The impulsive C2.8-class flare that occurred on 19 March 2023 (start/peak/end times: 02:12/02:15/02:19\,UTC) was located in the south-eastern quadrant of the solar disc, within the penumbral region of the leading sunspot of AR NOAA~13256. The principal characteristics of the AR and flare are summarised in Table~\ref{tab:main_params}. Despite its moderate GOES class and relatively simple photospheric magnetic configuration, the flare exhibited a complex spatio-temporal evolution.
\begin{table*}
\caption{$^*$Main properties of AR NOAA~13256 and the associated C2.8 flare on 19 March 2023.}
\label{tab:main_params}
\medskip
\begin{tabular}{c|c|c!{\vrule width 1.3pt}c|c|c}
\hline
AR Properties & Value & Instrument &
Flare Parameters & Value & Instrument \\
\hline

AR Number
& NOAA 13256
& SWPC/NOAA
& \multirow{3}{*}{Phases \,(UTC)}
& Start: 02:12
&
\\

\cline{1-3}

Coordinates
& S23E58
& \multirow{2}{*}{SDO/HMI}
&
& Peak: 02:15
&
\\

\cline{1-2}

Magnetic Class
& $\beta/\alpha$
&
&
& End: 02:19
& GOES-16
\\

\cline{1-3} \cline{4-5}

&
&
&
GOES Class
& C2.8
& (1--8\,\AA)
\\

\cline{4-5}

McIntosh Class
& Eho/Hsx
& SWPC/NOAA
&
Integrated Flux SXR
& $8.9\cdot10^{-4}$\,J$\cdot$m$^{-2}$
&
\\

\hline

Sunspot Area
& 250/60 MSH
& SDO/HMI
&
Optical Importance
& SF
& LEA ($H\alpha$) ($3^{**}$)
\\
\hline
\multicolumn{6}{l}{\footnotesize {$^*$The table is compiled using data from the SWPC NOAA catalogue (\href{https://www.ngdc.noaa.gov/stp/space-weather/swpc-products/daily_reports/solar_event_reports/}{https://www.ngdc.noaa.gov/stp/space-weather/})}} \\[-5pt]
\multicolumn{6}{l}{\footnotesize{$^{**}$~Observation quality index (3 out of 5; observing conditions were moderate).}}\\[-5pt]
\multicolumn{6}{l}{\footnotesize{MSH -- millionths of the solar hemisphere ($\approx 3.03\cdot10^{6}$\,km$^{2}$).}}\\[-5pt]
\end{tabular}\\
\end{table*}

According to the SOHO/LASCO CME catalogue, no coronal mass ejection (CME) was associated with this event, which is typical of confined flares of this class~\cite[]{Yashiro2004}. Nevertheless, an analysis of the extreme-ultraviolet~(EUV) dynamics observed by SDO/AIA~\cite[]{Koynash2024VAK} makes it possible to verify the absence or presence of small-scale eruptive activity (mini-eruptions), which is important for assessing the applicability of the standard CSHKP two-ribbon flare model to the development of the studied event.

This work employs multi-wavelength observations spanning the range from MW to HXR emission. The MW observations were obtained with the SOLARSPEL radio spectropolarimeter (with a cadence of $\approx1$\,s), the Siberian Radioheliograph (SRH; Badary Radio Astronomical Observatory, ISTP SB RAS; with a cadence of 3.52\,s), and the Nobeyama Radio Polarimeters (NoRP; with cadences of 0.1 and 1\,s). The localisation and dynamics of the MW emission sources were determined from SRH imaging observations at 4.0, 9.0, and 11.4\,GHz within the observed frequency range of 2.8--11.8\,GHz. The temporal evolution of the SXR flux was analysed using observations from the GOES-16/XRS instrument~\cite[]{Machol2020EXIS}. The fluxes measured in the 0.5--4\,\AA{} and 1--8\,\AA{} channels, with a cadence of 1\,s, were used for flare classification and for determining the temporal phases and evolution of the event. The flare morphology and magnetic-field configuration were investigated using observations from the Atmospheric Imaging Assembly (AIA;~\cite[]{Lemen2012AIA}) and the Helioseismic and Magnetic Imager (HMI;~\cite[]{Scherrer2012,Schou2012HMI}) onboard the Solar Dynamics Observatory (SDO). The AIA data were analysed with cadences of 12\,s and 24\,s in the EUV and UV channels, respectively, whereas the HMI magnetograms have a cadence of 720\,s. The SXR and HXR emission was studied using observations from the imaging spectrometers STIX onboard Solar Orbiter~\cite[]{Krucker2020STIX} and HXI onboard ASO-S~\cite[]{Zhang2019HXI}. High cadence time profiles and spectra were obtained from Fermi/GBM observations~\cite[]{Meegan2009Fermi}, with cadences of 256\,ms and 64\,ms during the triggered mode in the impulsive phase of the flare. The main characteristics of the instruments used in this study are summarised in Table~\ref{tab:dataInst}.
\begin{table*}[t]
\caption{Main characteristics of the instruments used for the analysis of flare SOL2023-03-19T02:14 (C2.8).}
\label{tab:dataInst}
\centering

\begin{tabular}{lcccl}
\hline
\multicolumn{5}{c}{\textbf{Non-imaging instruments}} \\
\hline
Instrument & Spectral Range & $\Delta t$ & & Purpose \\
\hline

SOLARSPEL &
3--24\,GHz &
$\sim$1\,s &
&
MW spectral analysis \\

NoRP &
1--80\,GHz &
0.1--1\,s &
&
MW temporal profiles and polarisation \\

GOES-16/XRS &
0.5--4 and 1--8\,\AA &
1\,s &
&
Flare classification and timing \\

Fermi/GBM &
4--150\,keV &
64--256\,ms &
&
HXR spectral analysis \\

\hline
\multicolumn{5}{c}{\textbf{Imaging instruments}} \\
\hline

Instrument &
Spectral Range &
$\Delta t$ &
$\theta_{\mathrm{res}}$ &
Purpose \\
\hline

SRH &
2.8--11.8\,GHz &
3.52\,s &
$\sim35''$ &
Localisation and dynamics of MW sources \\

SDO/AIA &
UV (1600\,\AA) &
24\,s &
$\sim1.5''$ &
Flare morphology and loop evolution \\

SDO/HMI &
6173\,\AA &
720\,s &
$\sim1''$ &
Magnetic-field analysis \\

Solar Orbiter/STIX &
4--150\,keV &
4\,s &
$\sim7''$ &
Localisation of HXR sources \\

ASO-S/HXI &
$\sim$14--200\,keV &
$\sim1$\,s &
$\sim9.3''$ &
Localisation of HXR sources \\
\hline
\end{tabular}
\end{table*}

The integrated MW flux was analysed using observations obtained with the Solar Radio Spectropolarimeter (SOLARSPEL; SOLAr Radio SPEctro poLarimeter; \citealt{Altyntsev2020}). SOLARSPEL is a centimetre-wavelength spectropolarimeter covering the frequency range 3--24\,GHz. The instrument consists of three prime-focus parabolic antennas operating over the 3--6, 6--12, and 12--24\,GHz sub-bands, respectively, whose signals are processed simultaneously by a digital receiver. Under normal operating conditions, SOLARSPEL records 48 frequency channels (16 channels per sub-band) with a bandwidth of 10\,MHz and have a a cadence of 0.96\,s. In the present study, only the 32 frequency channels covering the 3--12\,GHz range were used because the 12--24\,GHz antenna had not yet been commissioned at the time of the flare observations.

At the time of the event, SOLARSPEL had not yet been absolutely calibrated in physical units of flux density (solar flux units, sfu). Therefore, its data were cross-calibrated against observations of the quiet Sun obtained with the calibrated \textit{Nobeyama Radio Polarimeters} (NoRP) at 2, 3.75, 9.4, 17, and 35\,GHz. To estimate the reference fluxes at intermediate frequencies, including those below 3.75\,GHz, the NoRP spectrum was approximated by a second-order polynomial. The calibration procedure follows the principles of frequency-agile radio spectroscopy and implemented for instruments of this type~\cite[]{Zirin1991, Hurford1984Zirin}. This approach minimizes instrumental uncertainties in the absence of absolute calibration by referencing the entire observing band to accurately calibrated flux measurements. The uncertainties of the Stokes $I$ and $V$ parameters for each frequency channel were estimated as the root-mean-square (RMS) fluctuations measured during the pre-flare interval (02:07--02:14\,UTC), corresponding to the undisturbed solar atmosphere immediately before the flare onset. The uncertainty of the degree of circular polarization was calculated using standard error propagation for the ratio of two independent variables.

The calibration procedure consisted of the following steps:

\begin{enumerate}
\item The NoRP fluxes at the reference frequencies (2, 3.75, 9.4, 17, and 35\,GHz) were averaged over the selected quiet-Sun interval to obtain the reference spectrum.
\item For each of the 32 SOLARSPEL frequency channels in the 3--12\,GHz range, the reference flux density $S_{\mathrm{ref}}(\nu)$ was obtained by piecewise linear interpolation in logarithmic space, corresponding to a local power-law approximation of the form $S(\nu) \propto \nu^{\alpha}$ between adjacent NoRP reference frequencies.
\item The calibration coefficient for each frequency channel was then calculated as
 \begin{equation*}
        K(\nu) = \frac{S_{\text{ref}}(\nu)}{\langle P_{\text{obs}}(\nu) \rangle_{\text{quiet}}},
\end{equation*}
where $\langle P_{\text{obs}}(\nu) \rangle_{\text{quiet}}$ denotes the mean instrumental signal measured during the quiet-Sun reference interval.
\end{enumerate}

The use of five NoRP reference frequencies spanning the operational range of the spectropolarimeter SOLARSPEL provided an accurate reconstruction of the quiet-Sun spectrum and minimised interpolation uncertainties, particularly near the boundaries of the 3--12\,GHz frequency range.

In addition to the integrated MW flux measurements obtained with SOLARSPEL, data from SRH were also used. The SRH antenna systems have characteristics similar to those of the SOLARSPEL modules, including dish diameters of 3\,m and 2\,m for the corresponding frequency ranges and a bandwidth of 10\,MHz. Observations are performed using sequential frequency scanning with 16 frequency channels in each sub-band. As a result, the temporal cadence at a fixed frequency is 3.52\,s, whereas images obtained at neighbouring frequencies are offset in time by 0.22\,s. The acquisition of a full image at a single frequency in both circular polarisation channels (LCP and RCP) requires approximately $\approx 0.22$\,s.

The synthesis and calibration of SRH radio images were performed using software packages developed at the Institute of Solar--Terrestrial Physics, SB RAS, which implement self-calibration algorithms for individual scans together with parallel computing techniques~\cite[]{srhsynth, srhimages}. Antenna gain calibration was carried out at intervals of approximately $\approx 5$\,minutes throughout the observing period (00:00--10:00\,UTC) using a self-calibration procedure. To minimise instrumental effects, individual calibration solutions were further corrected through the extraction of a long-term daily phase trend using a differential evolution algorithm, compensation for phase noise, and exclusion of poorly calibrated baselines from the image synthesis process. The post-processing of antenna gain solutions provided a solar-disk positioning accuracy better than approximately one-half of the synthesized beam width.

The smoothed gain coefficients were interpolated onto the temporal grid of the flare observations and used to produce a sequence of MW images covering the 2.8--11.8\,GHz frequency range. The resulting image cube provides a maximum temporal resolution of 3.52\,s at each frequency. Images of $1024\times1024$ pixels, corresponding to a pixel size of $2.45''$, were synthesized without temporal averaging in order to preserve the rapid temporal variations of the flare emission. The dirty-beam sidelobes were removed using the CLEAN algorithm. At the highest observing frequency (11.8\,GHz), the full width at half maximum (FWHM) of the synthesized beam was approximately $\approx5''$. The beam shape was determined by both the orientation of the interferometric baselines with respect to the Sun at the time of the observations and the distribution of the available antenna pairs. Several baselines were either unavailable during the observations or deliberately excluded because of poor data quality and therefore did not contribute to the image synthesis.

For the analysis of QPPs in the SOLARSPEL data, Fourier spectral analysis was applied over the interval 02:13:40--02:16:20\,UTC. To improve the signal-to-noise ratio, frequency-averaged light curves were constructed for the 3--6\,GHz, 6--12\,GHz, and 3--12\,GHz bands. Prior to averaging, the flux density in each frequency channel was normalised to the 3\,GHz level. Specifically, the flux in the $n$-th channel was multiplied by a scaling factor $k=F_{3.0}/F_n$, where $F_{3.0}$ and $F_n$ denote the flux densities at 3\,GHz and at the corresponding channel frequency, respectively. A comparison of the Fourier power spectra obtained for different frequency ranges was used to assess the contribution of a narrow-band component observed after the main flare peak.

In addition to the Fourier analysis, a wavelet analysis was performed for the GOES/XRS time derivative, Fermi/GBM, and NoRP time series. The data were first interpolated onto a common temporal grid with a cadence of 1\,s over the interval 02:12--02:18\,UTC, which includes the flare impulsive phase together with the pre- and post-flare background. Long-term trends were removed by subtracting a smoothed profile obtained with a 21\,s moving window (other smoothing-window lengths were also tested). The detrended time series were then normalized to their peak values and additionally smoothed over 2\,s to suppress high-frequency fluctuations. We note that applying this additional 2\,s smoothing has little effect on the wavelet results: the same dominant periodicity is recovered, although the corresponding wavelet power is slightly reduced when no smoothing is applied. The processed time series were analysed using the Morlet mother wavelet implemented in the widely used \textit{Wavelet software provided by C. Torrence and G. Compo}~\cite[]{TorrenceCompo1998BAMS}. The statistical significance of the detected periodicities was evaluated by comparing the wavelet power spectrum with the corresponding red-noise background spectrum. A similar methodology has been adopted in previous studies~\cite[e.g.][]{Kupriyanova2010, Zimovets2023GiA, Zimovets2025CosRes}.

\section{RESULTS OF MULTI-WAVELENGTH OBSERVATIONS}
\subsection{Temporal Evolution of Flare Energy Release}
\begin{figure*}
\centering
\includegraphics[width=1.0\textwidth]{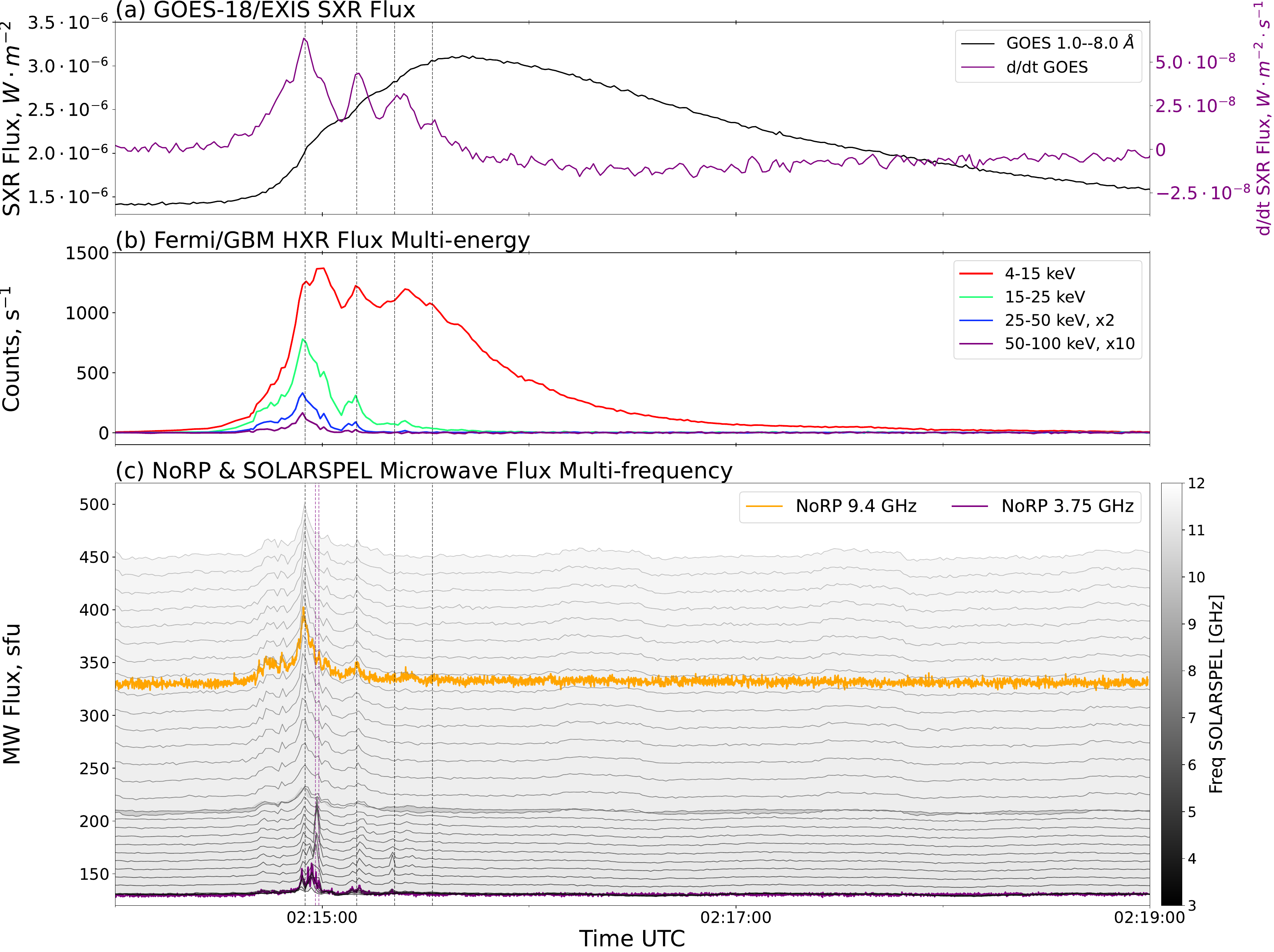}
\caption{Multi-wavelength observations of the C2.8 solar flare on 19 March 2023. Time profiles of the observed emission fluxes: (a) SXR flux recorded by GOES-16/XRS in the 1–8\,Å channel (black) together with its time derivative, $d/dt$ (purple); (b) HXR count rates measured by Fermi/GBM in the 4--15, 15--25, 25--50, and 50--100\,keV energy ranges; (c) MW flux density at selected frequencies: 3.75\,GHz (purple) and 9.4\,GHz (orange) from NoRP, together with SOLARSPEL observations in the 3–12\,GHz frequency range.
} 
\label{fig1}
\end{figure*}

The temporal evolution of the flare emission in different wavelength ranges is presented in Fig.~\ref{fig1}. Panel (a) shows the SXR flux and its time derivative obtained from GOES-16/XRS observations in the 1--8\,\AA{} channel. Panel (b) displays the HXR time profile derived from Fermi/GBM observations, while panel (c) presents the MW flux measured by SOLARSPEL and NoRP. All data are shown on a common UTC time scale. The duration of the impulsive phase was approximately 1\,min.

The HXR time profile and the derivative of the SXR flux (Fig.~\ref{fig1}) clearly reveal at least four consecutive peaks (pulsations). Cross-correlation analysis did not reveal any statistically significant delays between the emission time profiles observed in different spectral ranges on timescales exceeding 1\,s. The close temporal correspondence between the impulsive components observed in the MW and HXR ranges and in the SXR derivative is consistent with the Neupert effect~\cite[]{Neupert1968} operating for the individual peaks, indicating a series of electron injections into the flare loops.

A notable feature of this event is the apparent decrease in the time interval between successive peaks. According to the GOES/XRS SXR derivative, the intervals between the first and second peaks, second and third peaks, and the third-to-fourth and fourth-to-fifth peaks are approximately $P_{1-2}\approx15$\,s, $P_{2-3}\approx13$\,s, and $P_{3-4},P_{4-5}\approx9$\,s, respectively. The mean period between the peaks is estimated to be approximately $11.5\pm3.0$\,s.

\begin{figure*}
\centering
\includegraphics[width=0.65\textwidth]{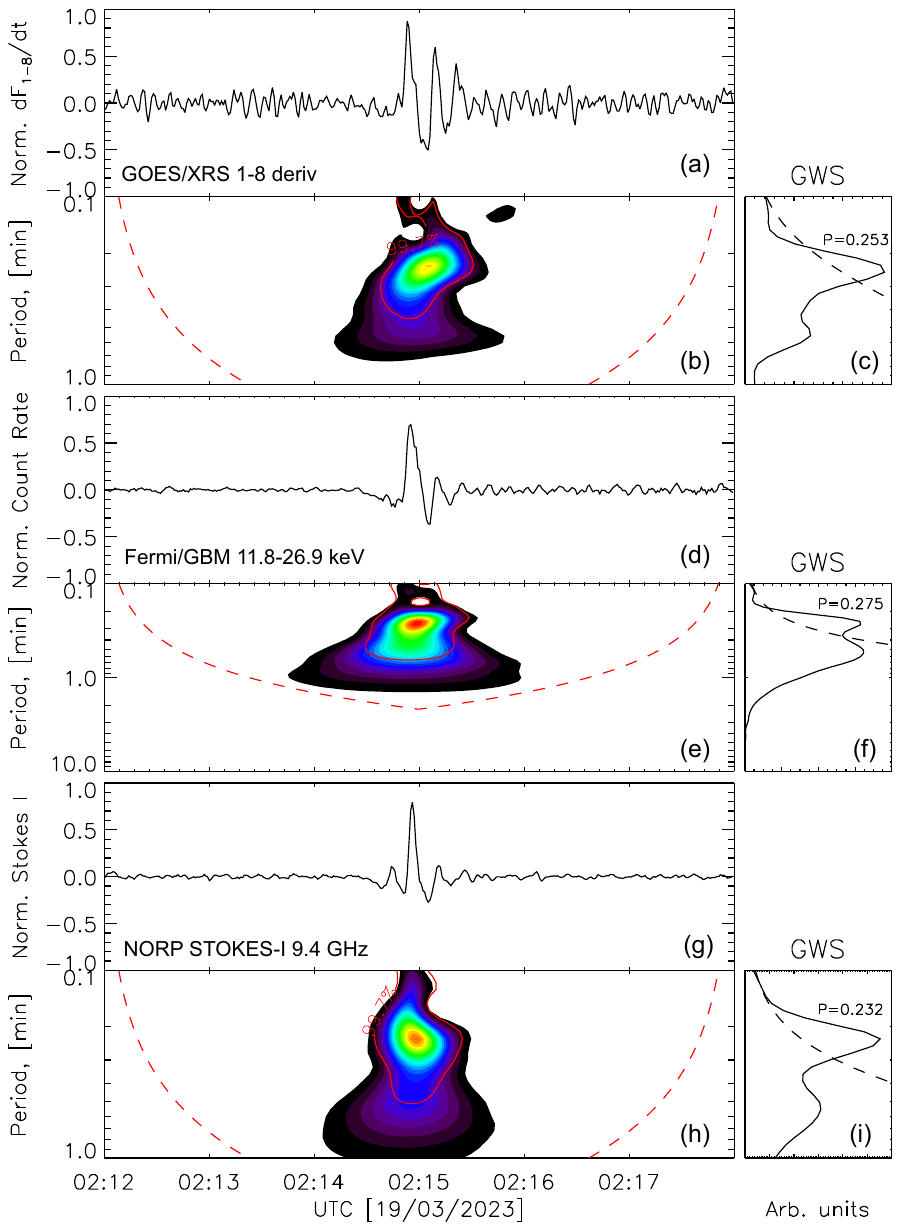}
\caption{Visualization of wavelet analysis of the temporal profiles of the flux time derivative in the GOES/XRS 1–8\,\AA{} channel (a--c), of the count rate in the Fermi/GBM 11.8–26.9\,keV channel (d--f), and NoRP Stokes~$I$ at 9.4\,GHz. The panels (a, d, g) show the prepared temporal profiles with the subtracted trend and normalized. The panels (b, e, h) show the wavelet spectra, where the solid curve shows the 99.7\% confidence levels, and the dashed curve shows the cone of influence. On the panels (c, f, i) the solid line shows the global wavelet spectrum (GWS) and
the dashed line is the spectrum of the red noise model. The values of significant spectral peaks are indicated (in minutes).} 
\label{wavelet}
\end{figure*}

The shortening of the characteristic period from $\approx15$\,s to $\approx9$\,s is also visible in the wavelet spectrum of the GOES/XRS SXR flux derivative (Fig.\,\ref{wavelet}\,b). The global wavelet power spectra of the detrended and normalised signals (Fig.\,\ref{wavelet}\,c,f,i) reveal a statistically significant peak with a period of $P_{\mathrm{QPP}}\approx15.2\pm1.3$\,s. In particular, the periods corresponding to the maximum spectral power are $P\approx15.2$\,s for the SXR derivative, $P\approx16.5$\,s for the HXR emission in the 11.8--26.9\,keV range, and $P\approx13.9$\,s for the NoRP 9.4\,GHz MW emission. Within the uncertainties, these values are consistent with the period of $P=14\pm1$\,s obtained from the Fourier analysis of the SOLARSPEL MW time profiles.

We therefore conclude that QPPs were present during the impulsive phase of the flare in the MW and HXR emissions, as well as in the time derivative of the SXR flux. In contrast, no significant QPP signatures were detected directly in the SXR flux profiles ($\lesssim10$\,keV). Additional characteristics of the observed QPPs include the small number of pulses, their rapid decay, and the distinctly non-sinusoidal waveform characterised by sharp peaks.

\begin{figure*}
\centering
\includegraphics[width=1.0\textwidth]{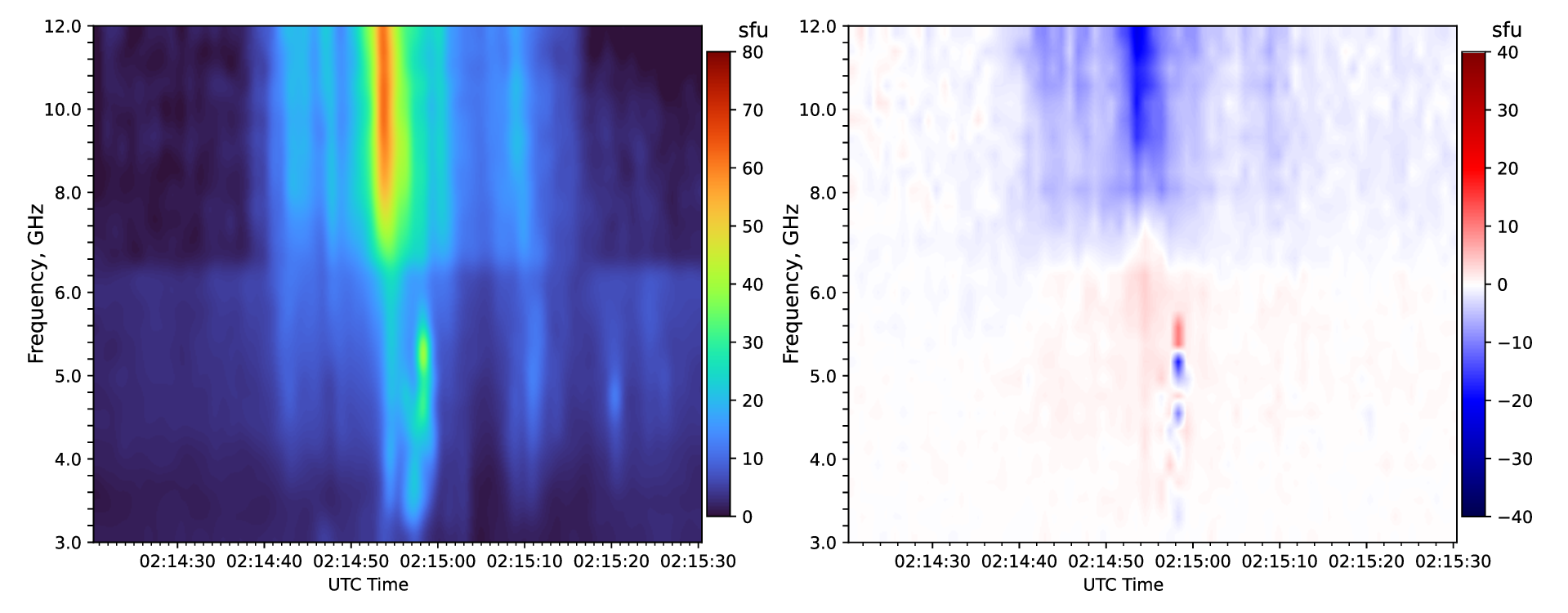}
\caption{Dynamic spectra of the MW emission recorded by SOLARSPEL.
The left panel shows the Stokes~$I$ parameter (total intensity), while the right panel shows the Stokes~$V$ parameter (circular polarisation).
} 
\label{DYNspect}
\end{figure*}

\begin{figure*}
\centering
\includegraphics[width=1.0\textwidth]{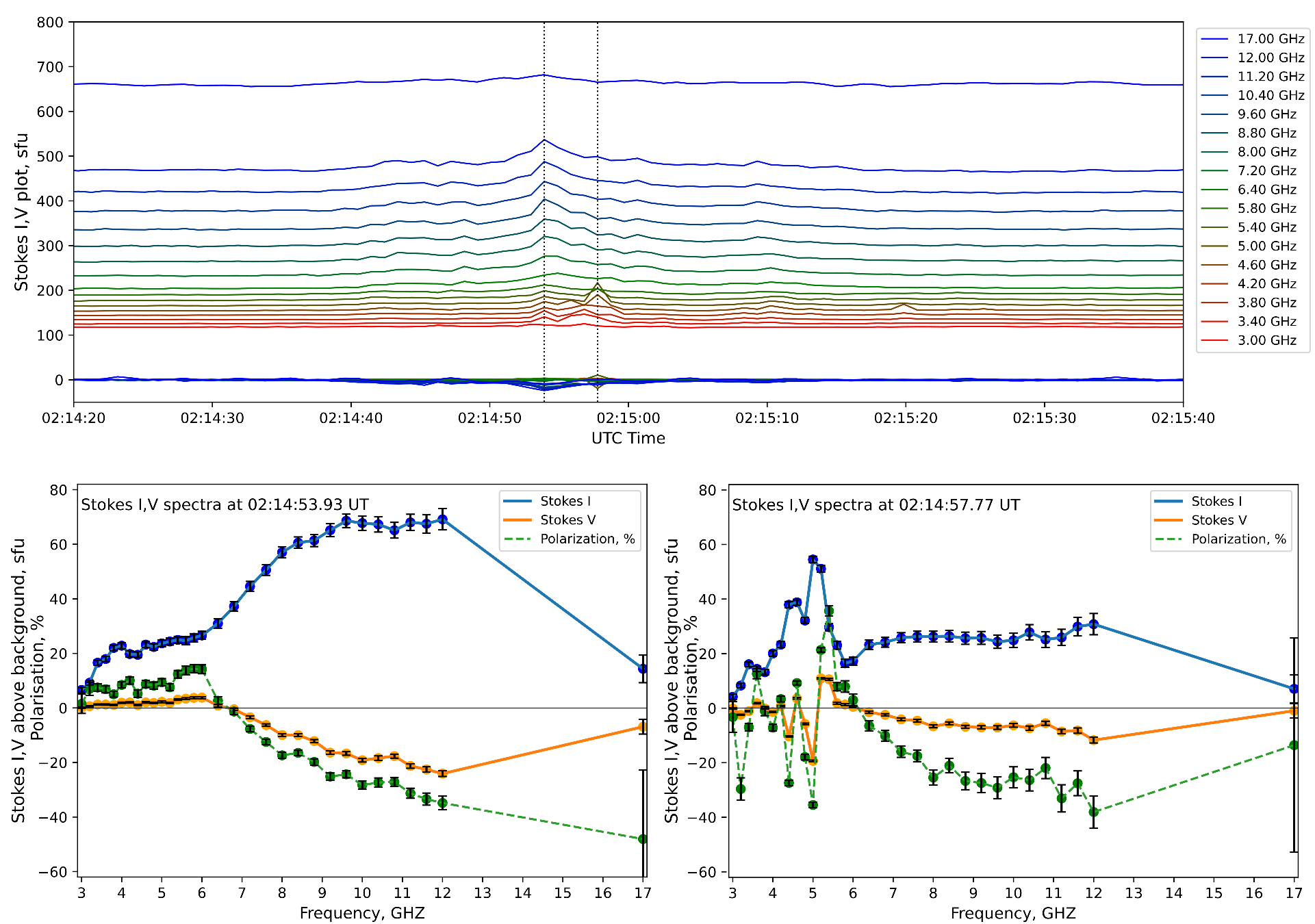}
\caption{MW spectra obtained with SOLARSPEL and NoRP.\protect\\
{\it Top panel}: Time profiles of the absolute values of the Stokes~$I$ and $V$ parameters for selected SOLARSPEL frequency channels in the 3–12\,GHz range, supplemented by the NoRP 17\,GHz time profile. The vertical dashed lines indicate the times at which the combined MW spectra were constructed.\protect\\
{\it Bottom panel}: Combined spectra of the relative Stokes $I$ and $V$ parameters and the degree of circular polarisation derived from the SOLARSPEL + NoRP observations for the flare peak (left) and for the subsequent short-duration narrowband burst (right).
} 
\label{MWspect}
\end{figure*}

The dynamic spectra of the MW emission are presented in Fig.~\ref{DYNspect} using SOLARSPEL observations. A noteworthy feature is a narrowband coherent burst observed in the 4.0--4.6\,GHz frequency range (Fig.~\ref{MWspect}), marked by two purple dashed lines in Fig.~\ref{fig1}\,c. This burst occurred approximately 4\,s after the flare maximum.

The flare also exhibited a reversal of the circular polarisation sign (Stokes $V$) near $\sim6.7$\,GHz (Fig.~\ref{MWspect}). Independent observations obtained with SOLARSPEL, NoRP, and SRH confirm the reliability of this effect. At frequencies below 6\,GHz, the Stokes $V$ component is positive, whereas at higher frequencies the circular polarisation becomes negative.

A detailed investigation of the narrowband coherent burst and the polarisation reversal is beyond the scope of the present study.

\subsection{Spatial Structure of the Flare Energy Release}

To investigate the spatial structure of the flare, UV, X-ray, and radio observations were analysed. The imaging data show that the flare developed in AR NOAA~13256 in the vicinity of a large sunspot (approximately $\approx 30''$ in diameter) and was associated with a system of magnetic loops located near the penumbra of the leading sunspot. Two flare ribbons were identified at the footpoints of the loop system, with characteristic lengths of approximately $15''$ and $30''$ for the eastern and western ribbons, respectively. The shorter eastern ribbon was partially embedded within the sunspot penumbra. Both ribbons exhibited a fragmented structure consisting of several localised brightenings with characteristic sizes of $3''$--$7''$. In total, approximately 8--10 brightenings were identified within the two ribbons, corresponding to roughly four brightenings per ribbon, comparable to the number of observed QPP peaks.

\begin{figure*}
\centering
\includegraphics[width=1.0\textwidth]{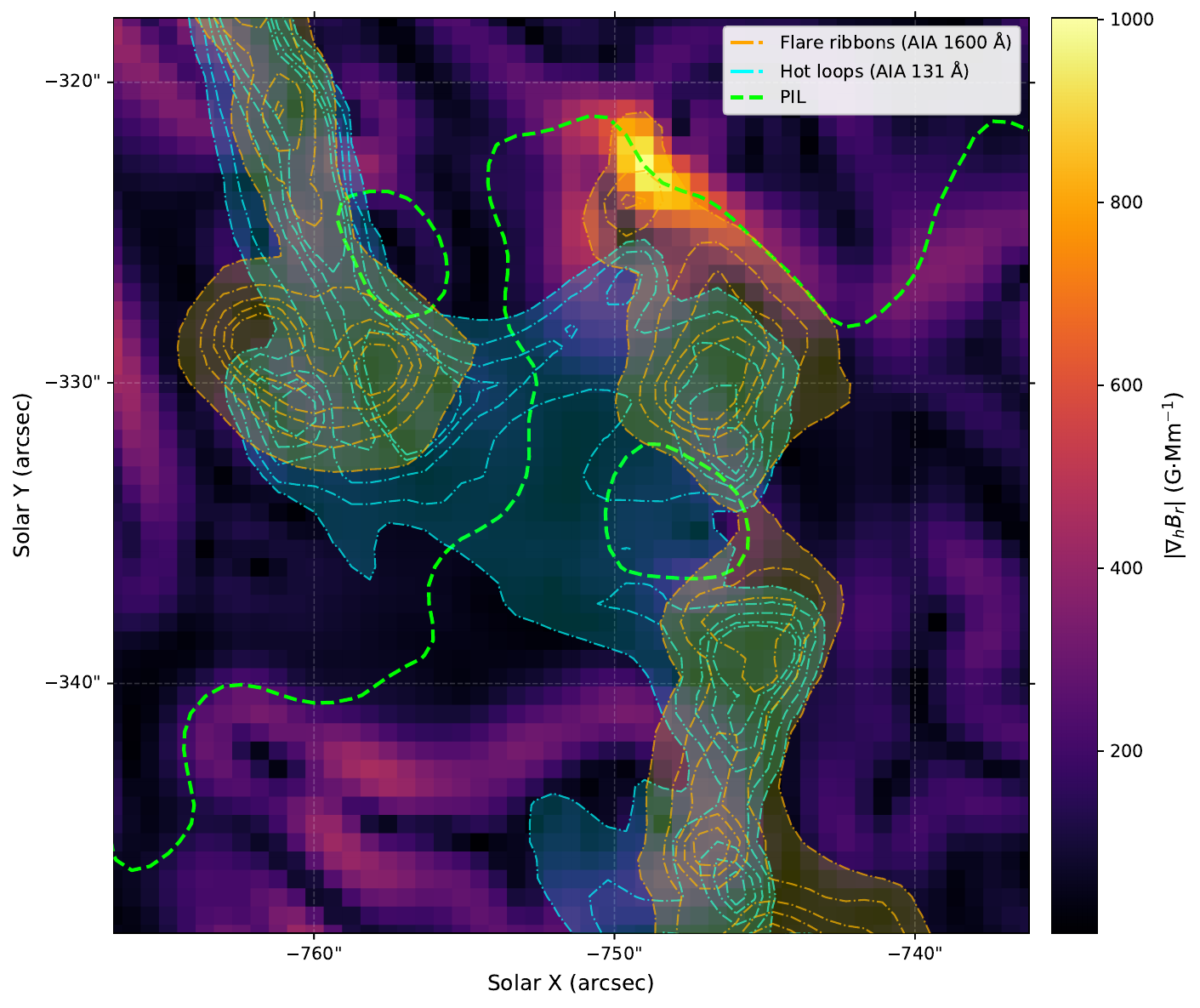}
\caption{
Map of the horizontal gradient of the radial photospheric magnetic field, $|\nabla_h B_r|$, derived from SDO/HMI observations. The colour scale represents the magnitude of the horizontal magnetic-field gradient in units of
G\,Mm$^{-1}$. The green dashed curve denotes the PIL.
Orange contours and shaded regions show the flare ribbons observed in the SDO/AIA 1600\,\AA{} channel, while cyan contours and shading indicate the hot coronal loops observed in AIA 131\,\AA{}. The flare ribbons are located on
opposite sides of the PIL in the vicinity of the strongest magnetic-field gradients, whereas the hot coronal loops connect the conjugate ribbon sources. The close spatial correspondence between these structures indicates that the
primary energy release occurred within a compact magnetic system rooted near the PIL.
}
\label{grad_BR}
\end{figure*}

Figure~\ref{grad_BR} shows the distribution of the horizontal gradient of the
radial photospheric magnetic field, $|\nabla_h B_r|$, derived from SDO/HMI
observations. The PIL passes through the central part of the analysed region and coincides with a localized area where the magnetic field exhibits the largest horizontal gradients, reaching $|\nabla_h B_r|\approx1\,\mathrm{kG\,Mm^{-1}}$.

The flare ribbons observed in the AIA\,1600\,\AA{} channel are situated on
opposite sides of the PIL and are concentrated within this high-gradient
region. At the same time, the hot coronal loops observed in AIA\,131\,\AA{}
connect the conjugate ribbon sources, outlining the magnetic structures
involved in the flare energy release.

Such a spatial configuration strongly suggests that the primary magnetic reconnection was initiated within a compact magnetic system located near the
PIL, where the magnetic field changes most rapidly over short spatial scales.
The coincidence of the flare ribbons, hot loops, and the region of enhanced
$|\nabla_h B_r|$ supports the interpretation that the initial energy release
occurred in low-lying sheared magnetic loops rooted close to the PIL.

The observed morphology is consistent with the current understanding of flare development in regions of strong magnetic-field gradients, where current sheets can form and magnetic reconnection can efficiently proceed. The maximum horizontal magnetic-field gradient reaches $\nabla_h B_r \approx 1000$,G,Mm$^{-1}$ and is located at helioprojective coordinates ($-749''$, $-323''$), near the northern footpoints of the flare loops associated with the western ribbon. This location subsequently coincided with the positions of the MW, EUV/UV, SXR, and HXR emission sources.

The close spatial association between regions of enhanced magnetic-field gradient, flare ribbons, and hot coronal loops suggests that particle acceleration was concentrated near the PIL. To test this hypothesis, we analysed the spatial distribution of MW and hard X-ray sources, which serve as direct tracers of accelerated electrons.

\begin{figure*}
\centering
\includegraphics[width=1.0\textwidth]{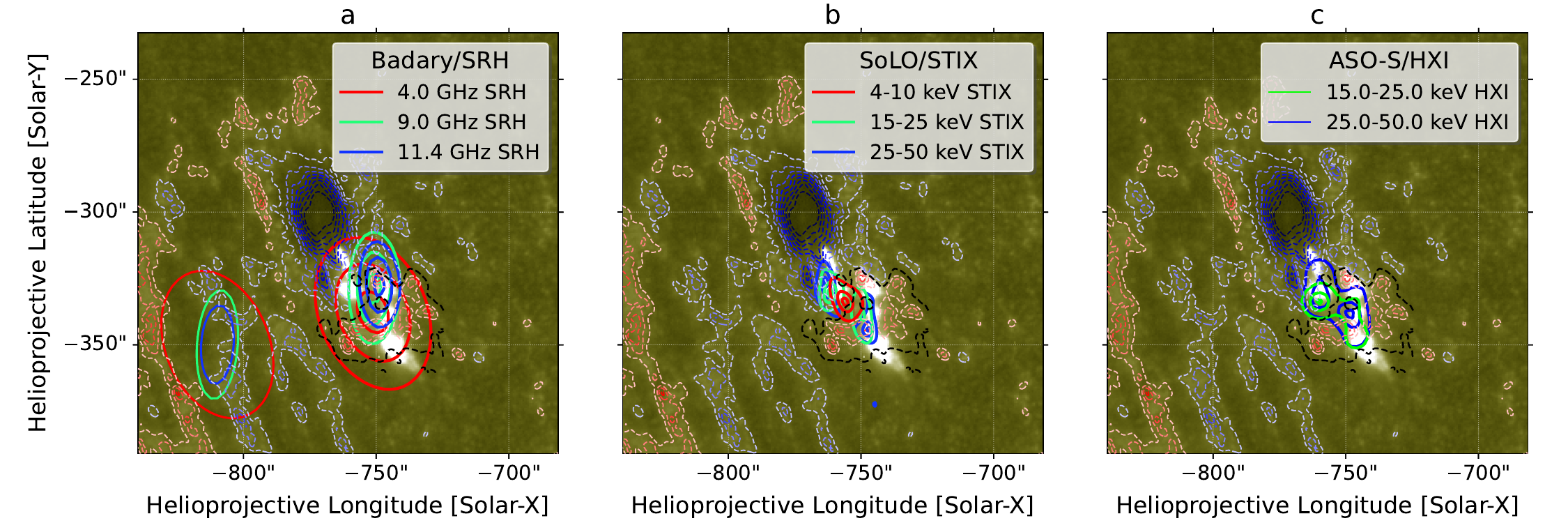}
\caption{Co-aligned UV, MW, and X-ray observations of the flare at its peak (02:14:55,UTC). The background image shows the SDO/AIA 1600\,\AA{} channel. Coordinates are given in arcseconds relative to the solar-disc centre. Contours correspond to intensity levels of 30\%, 70\%, and 95\% of the maximum intensity in each observing channel, and the black curve marks the PIL ($B_r=0$).
(a) Badary/SRH: MW emission contours at 4.0, 9.0, and 11.4\,GHz. The corresponding synthesized-beam FWHM is shown in the lower-left corner. (b) Solar Orbiter/STIX: X-ray emission contours reconstructed from observations integrated over 02:08:52--02:13:16\,UTC in the energy ranges 4--10, 15--25, and 25--50\,keV. (c) ASO-S/HXI: hard X-ray emission contours reconstructed from observations integrated over 02:14:36--02:15:02\,UTC in the energy ranges 15--25 and 25--50\,keV.}
\label{SRH_AIA}
\end{figure*}

The locations of the MW and HXR sources were determined using imaging observations from the Siberian Radioheliograph (SRH) and the STIX and HXI X-ray telescopes. The resulting MW and X-ray contours are overlaid on an AIA 1600\,\AA{} image obtained near the flare peak and are shown in Fig.~\ref{SRH_AIA}.

Analysis of the SRH images revealed a single compact MW source at all available frequencies between 3 and 12\,GHz (Fig.~\ref{SRH_AIA}a). The observed source size is comparable to the full width at half maximum (FWHM) of the SRH synthesized beam at the corresponding frequencies, indicating that the source remained unresolved. This implies that the intrinsic size of the MW emitting region is smaller than the spatial resolution of the SRH.

The HXR sources identified from STIX and HXI observations spatially coincide with the flare ribbons (Fig.~\ref{SRH_AIA}b,c). STIX observations reveal a single SXR source in the 4--10\,keV range, with a characteristic size of approximately $15''$ measured at the 50\% intensity contour. This source is projected between the flare ribbons and is interpreted as hot plasma located near the apex of the flare loop system. In the non-thermal HXR range (15--50\,keV), both STIX and HXI observations reveal two groups of compact sources associated with the ribbons located on opposite sides of the PIL. The extent of these HXR source groups is considerably smaller than the lengths of the UV ribbons, indicating that the most energetic processes were confined to a more compact region than the area exhibiting enhanced UV emission. The imaging data further suggest that the HXR footpoint sources and the thermal SXR source belong to the same magnetic loop system. Since the thermal source is clearly located between two non-thermal HXR source groups and the AR was observed relatively close to the limb, the flare loop hosting the accelerated particles was likely low-lying, with a height smaller than the separation between the opposite flare footpoints, i.e. $\lesssim 5$--$10$\,Mm.

The location of the MW and SXR sources between the two flare ribbons and HXR footpoint sources (Fig.~\ref{SRH_AIA}\,a, b) indicates that the MW emission most likely originated from the same system of low-lying flare loops.

The peak brightness temperature near the maximum of the MW spectrum ($\sim10$\,GHz) reached approximately $T_b\sim10^7$\,K. Since the source is compact and remains unresolved by the SRH synthesized beam, the intrinsic brightness temperature must be substantially higher. Assuming a loop width of approximately 3\,Mm, correcting for beam dilution using the ratio of the beam and source solid angles, $T_b,\Omega_{\rm beam}/\Omega_{\rm source}$, yields a lower limit on the characteristic energy of the emitting electrons of approximately $\approx10$\,keV. The high brightness temperature, compact source size, spatial association with the HXR source, and impulsive temporal behaviour all support a non-thermal gyrosynchrotron origin of the observed MW emission.

\subsection{Dynamics of the MW Sources}
\begin{figure*}
\centering
\includegraphics[width=1.0\textwidth]{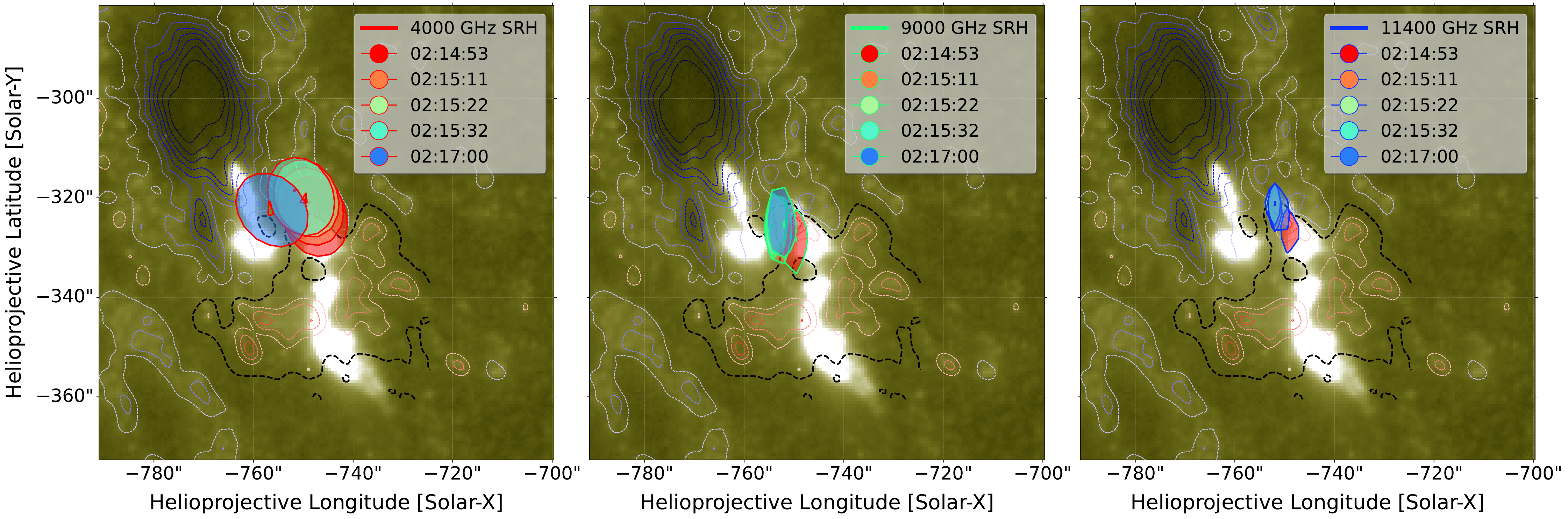}
\caption{Temporal evolution of the compact MW source during the impulsive phase of the flare at three observing frequencies: (a) 4.0\,GHz, (b) 9.0\,GHz, and (c) 11.4\,GHz. The background image in all panels is the SDO/AIA 1600\,\AA{} map. The coloured filled contours show the brightest ($95\%$--$99.999\%$ of the peak intensity) part of the background-subtracted SRH source at five successive times (see legends). The corresponding contour outlines are drawn in red (4.0\,GHz), green (9.0\,GHz), and blue (11.4\,GHz). Dashed coloured contours indicate the radial magnetic field component $\nabla_h B_r$ derived from SDO/HMI, while the black dashed line marks the polarity inversion line (PIL). The compact MW source exhibits systematic positional changes throughout the impulsive phase, with the largest displacement occurring close to the flare maximum. The source motion is approximately parallel to the flare ribbons and remains confined to the vicinity of the PIL and the region of the strongest horizontal gradient of the radial magnetic field.}
\label{AIA_SRH_timeevo}
\end{figure*}

Detailed evolution of the spatial morphology of the flare MW source is illustrated by co-aligned maps at 4.0, 9.0, and 11.4\,GHz obtained at successive times during the impulsive phase of the flare (Fig.\,\ref{AIA_SRH_timeevo}). At all three frequencies, the bright flare component exhibits systematic displacements with respect to the stable sunspot-associated source. The largest positional shifts occur near the flare maximum, coinciding with the most intense phase of energy release. The apparent motion of the MW source in the plane of the sky is approximately parallel to the flare ribbons. The sequence of the 95\% intensity contours shows that the brightest MW emission originates from a compact region located close to the chromospheric flare ribbons and the PIL, in the vicinity of the region with the strongest horizontal gradient of the radial magnetic field, $|\nabla_h B_r|$. Although the measured displacements are small compared with the SRH synthesized beam, the positions of the brightness maxima evolve systematically with time. This behaviour suggests that the observed motion of the MW source is physically associated with the evolution of the flare energy-release region and the underlying particle-acceleration process.

To quantify the spatial evolution of the MW source, a sequence of SRH images was analysed at each observing frequency. For every image, the source position was defined as the brightness-weighted centroid calculated within the 95\% intensity contour enclosing the compact flare-related emission. The centroid coordinates were measured in the helioprojective coordinate system and tracked throughout the impulsive phase of the flare. Apparent source velocities were estimated from finite differences between successive centroid positions divided by the corresponding time interval between SRH images. The uncertainties of the centroid coordinates were evaluated from the image pixel scale and the signal-to-noise ratio of the source and were subsequently propagated to the velocity estimates. Although the measured displacements are smaller than the FWHM of the synthesized beam, the centroid position can be determined with substantially higher precision owing to the high signal-to-noise ratio of the flare source.

\begin{figure*}[htbp]
  \centering
  \includegraphics[width=0.45\textwidth]{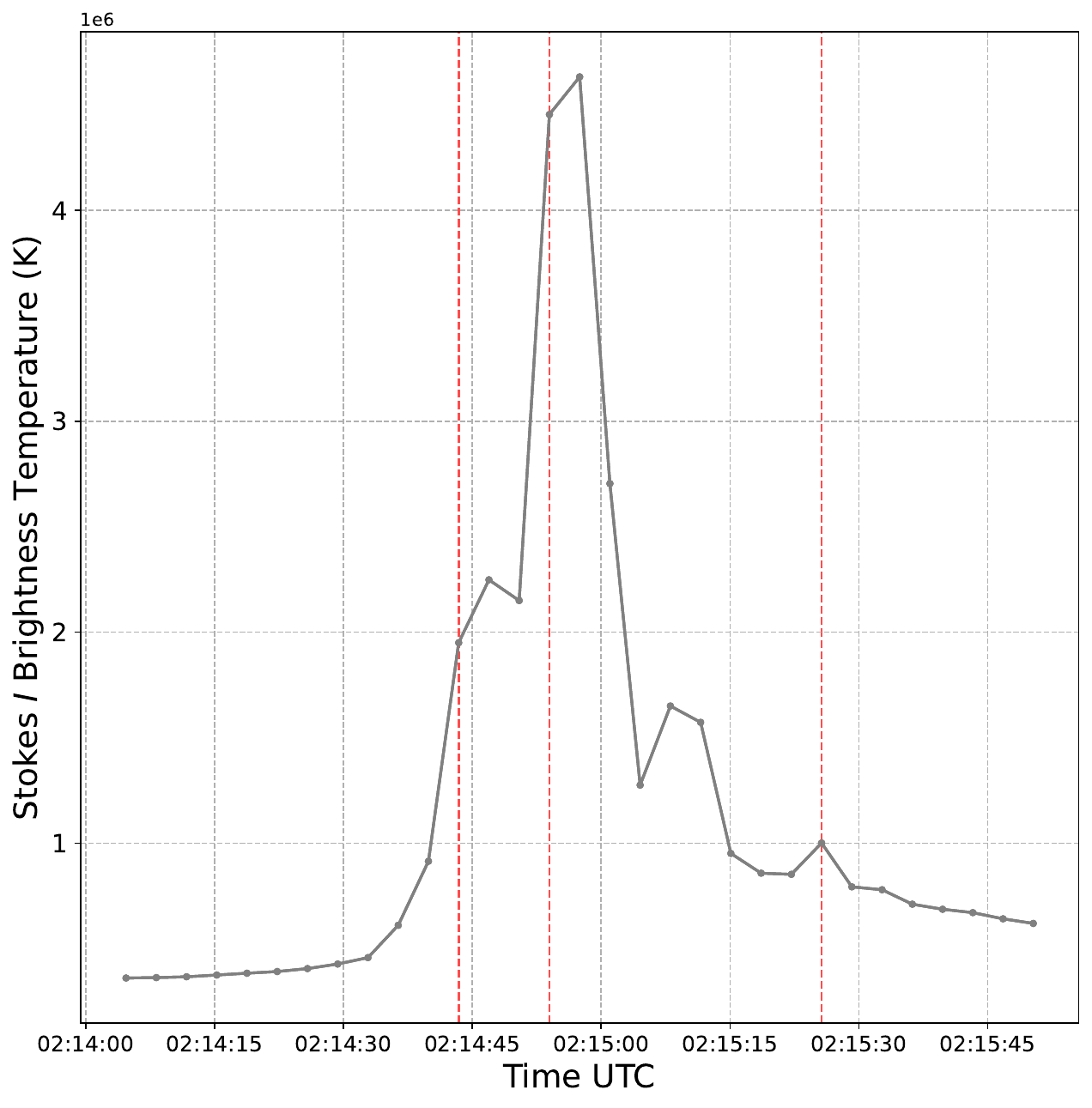}\hfill
  \includegraphics[width=0.5\textwidth]{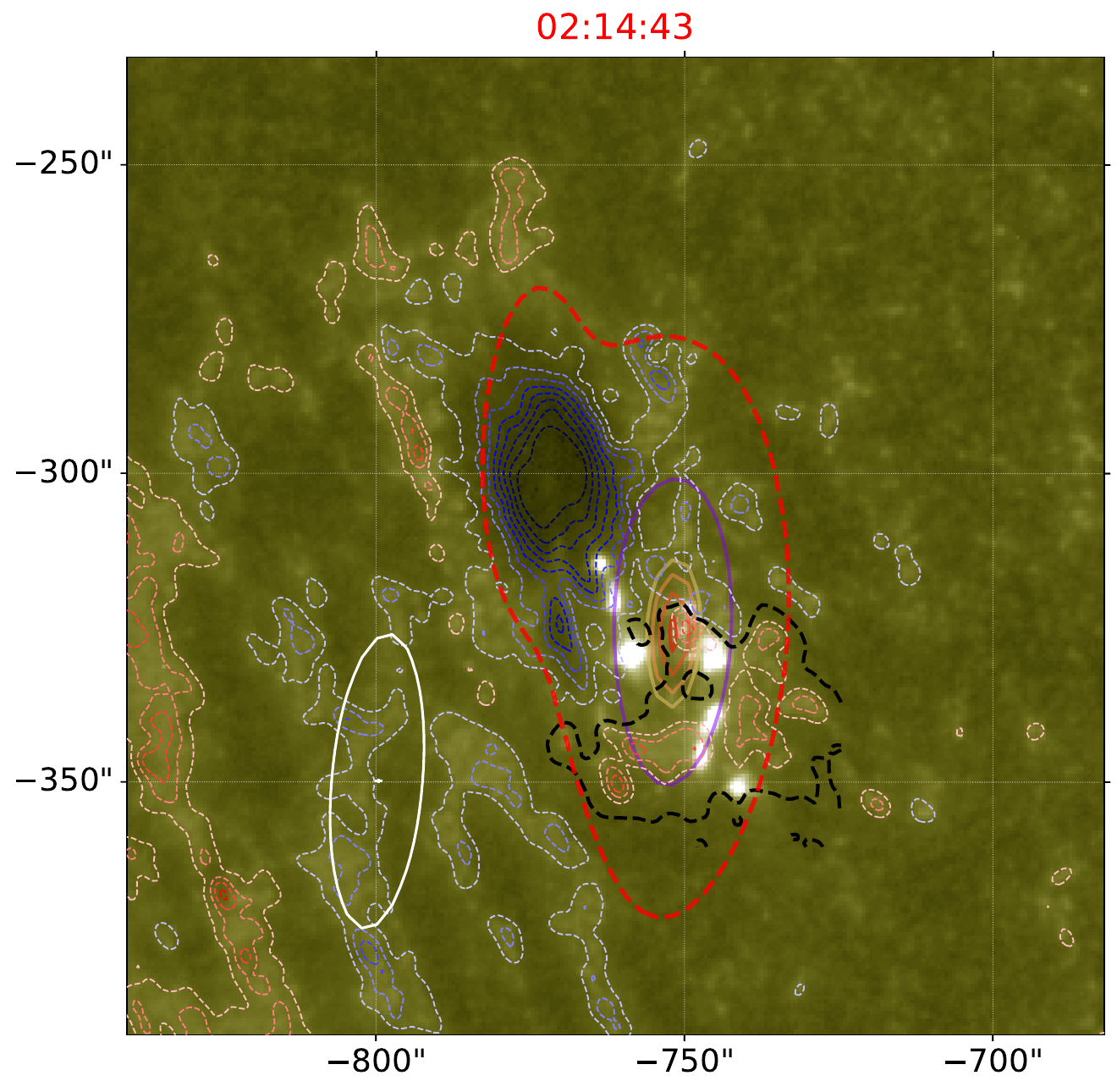}  
  \vspace{0.5cm}
  \includegraphics[width=0.5\textwidth]{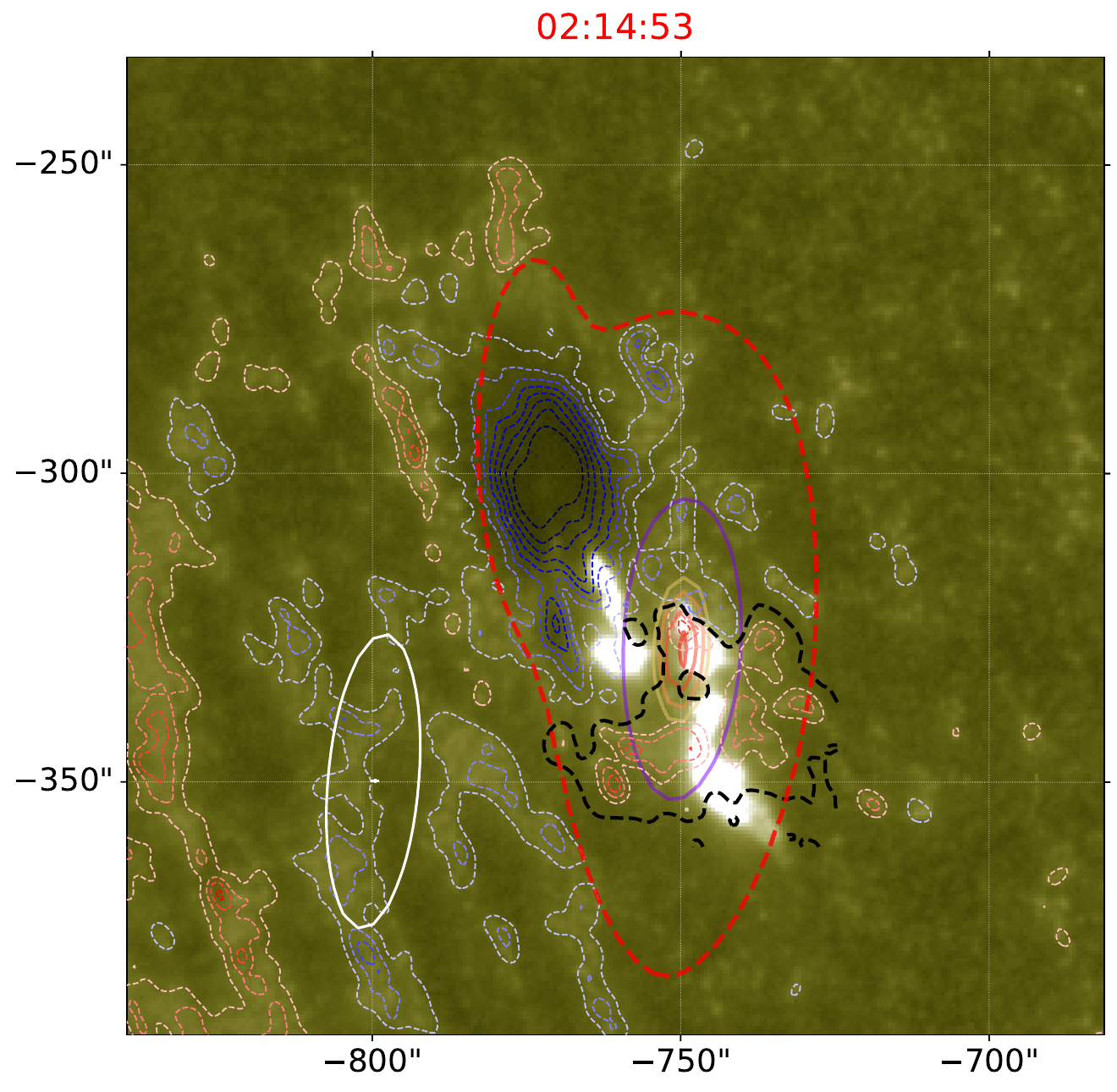}\hfill
  \includegraphics[width=0.5\textwidth]{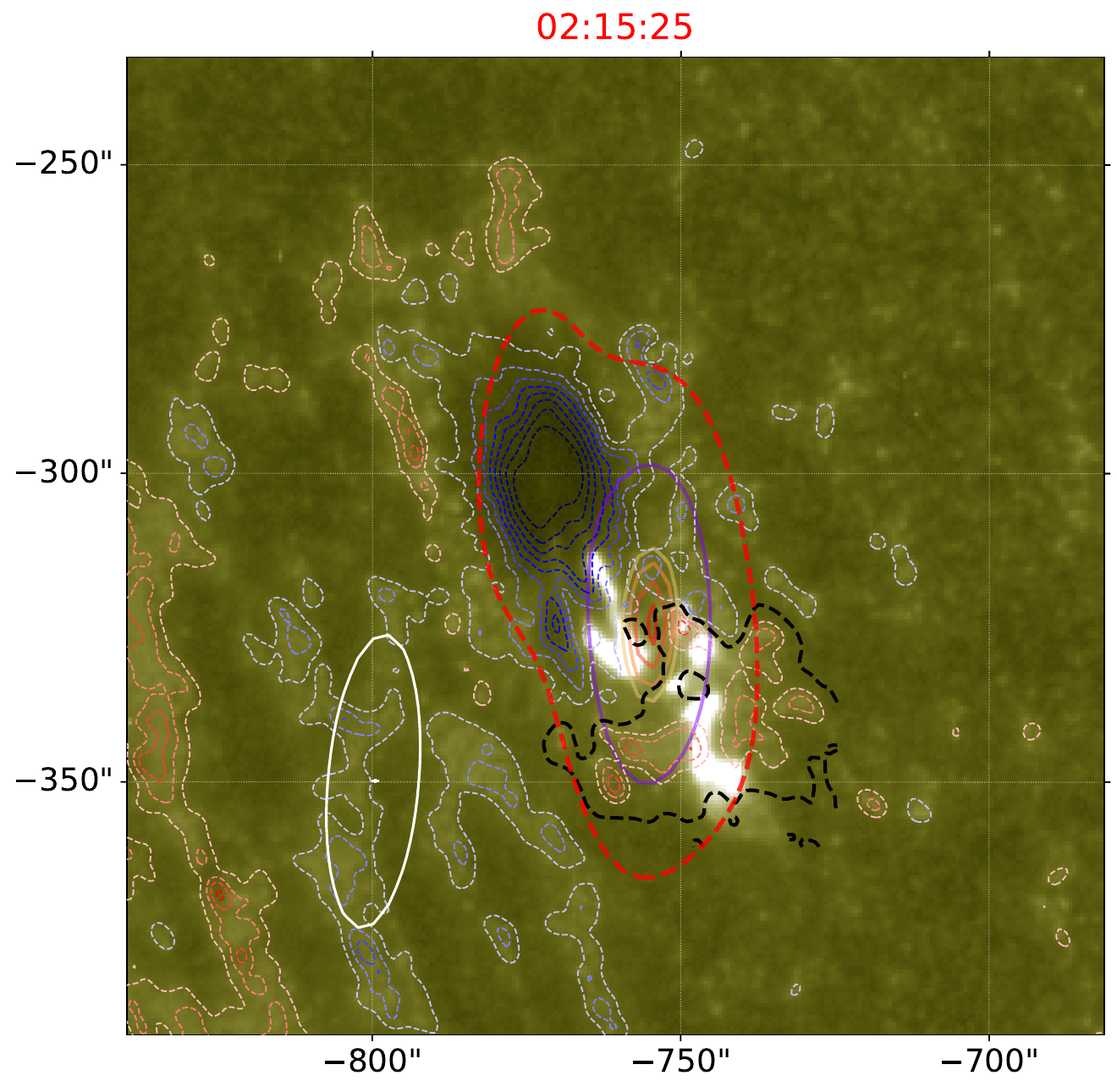}
  \caption{Evolution of the compact MW source at 9\,GHz during the impulsive phase. The upper-left panel shows the SRH Stokes~$I$ light curve; red dashed lines mark the three times displayed in the remaining panels. The upper-right, lower-left, and lower-right panels show co-aligned SDO/AIA~1600\,\AA{} images at 02:14:43, 02:14:53, and 02:15:25\,UTC, respectively, with overlaid SRH contours. The red dashed contour indicates the pre-flare sunspot-associated MW source (50\% level), coloured contours show the flare-related MW emission (50--99\% of the peak brightness temperature), white contours denote the FWHM of the SRH beam, the thick black dashed curve marks the PIL, coloured dashed contours correspond to the radial magnetic field~$B_r$, and the white cross indicates the estimated positional uncertainty of the MW source.}
  \label{AIA_SRH_timeevo_inline}
\end{figure*}

To investigate the spatial evolution of the MW sources in greater detail, a sequence of SRH radio maps at 9\,GHz, corresponding approximately to the spectral peak frequency of the flare, was analysed. Figure\,\,\ref{AIA_SRH_timeevo_inline} presents three representative moments corresponding to the pre-peak, peak, and post-peak stages of the MW emission. The image sequence reveals the presence of a relatively stable sunspot-associated radio source together with a brighter flare component whose position changes during the impulsive phase. The spatial coincidence of the sunspot radio source with the sunspot observed in the AIA~1600\,\AA{} channel provides an independent verification of the accuracy of the co-alignment between the SRH and SDO/AIA datasets.

Despite the compact nature of the flare source, its centroid exhibits a systematic displacement throughout the impulsive phase. The amplitude of the identified shifts is comparable to the image pixel size ($\approx2.45''$), which is substantially smaller than the SRH beam width. The observed source size remains close to the FWHM of the instrumental beam during the entire impulsive phase, indicating the absence of significant expansion of the MW-emitting region. This result is consistent with the confined nature of the event, during which neither the formation of an extended flare arcade nor substantial growth of the magnetic loops containing the MW-emitting accelerated electrons is observed. Consequently, the subsequent analysis focuses on the temporal evolution of the MW source centroid as a possible manifestation of non-stationary energy release and QPPs.

\begin{figure*}[htbp]
  \centering
  \includegraphics[width=1.0\textwidth]{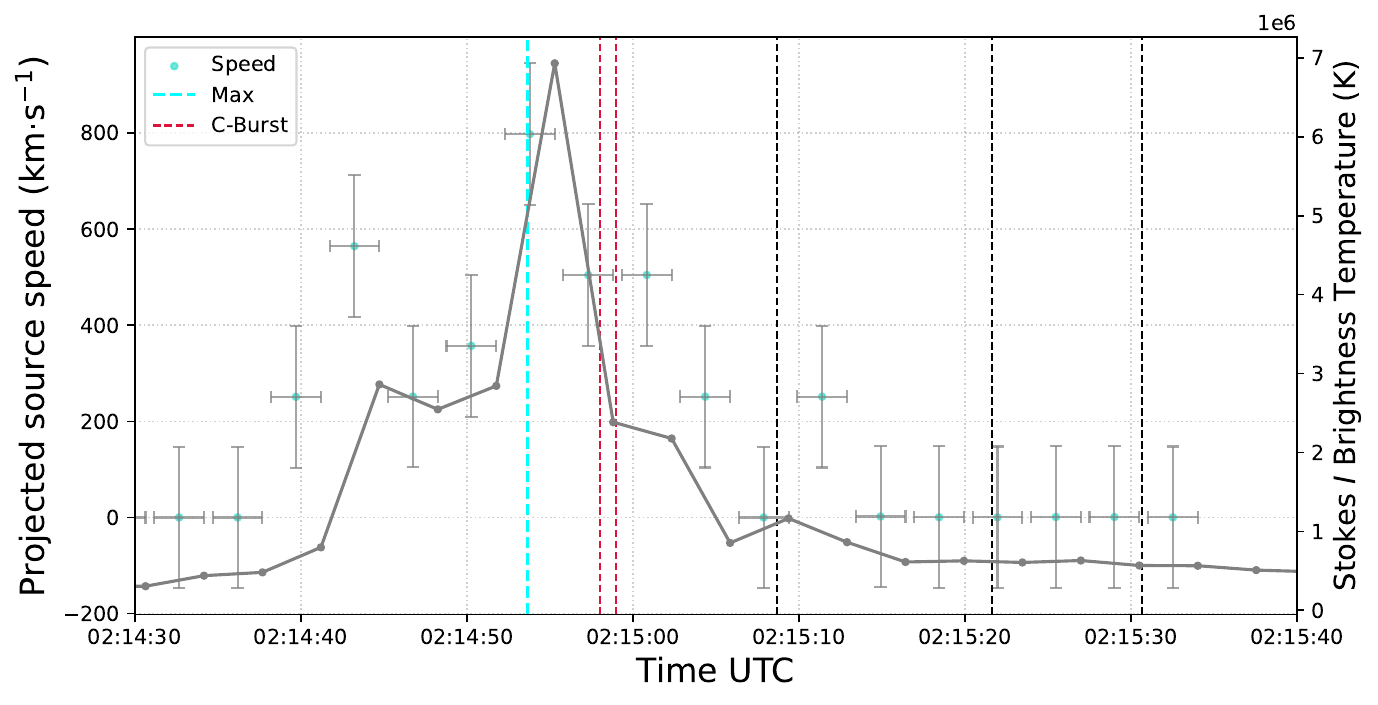}  
  \vspace{0.5cm} 
  \includegraphics[width=0.45\textwidth]{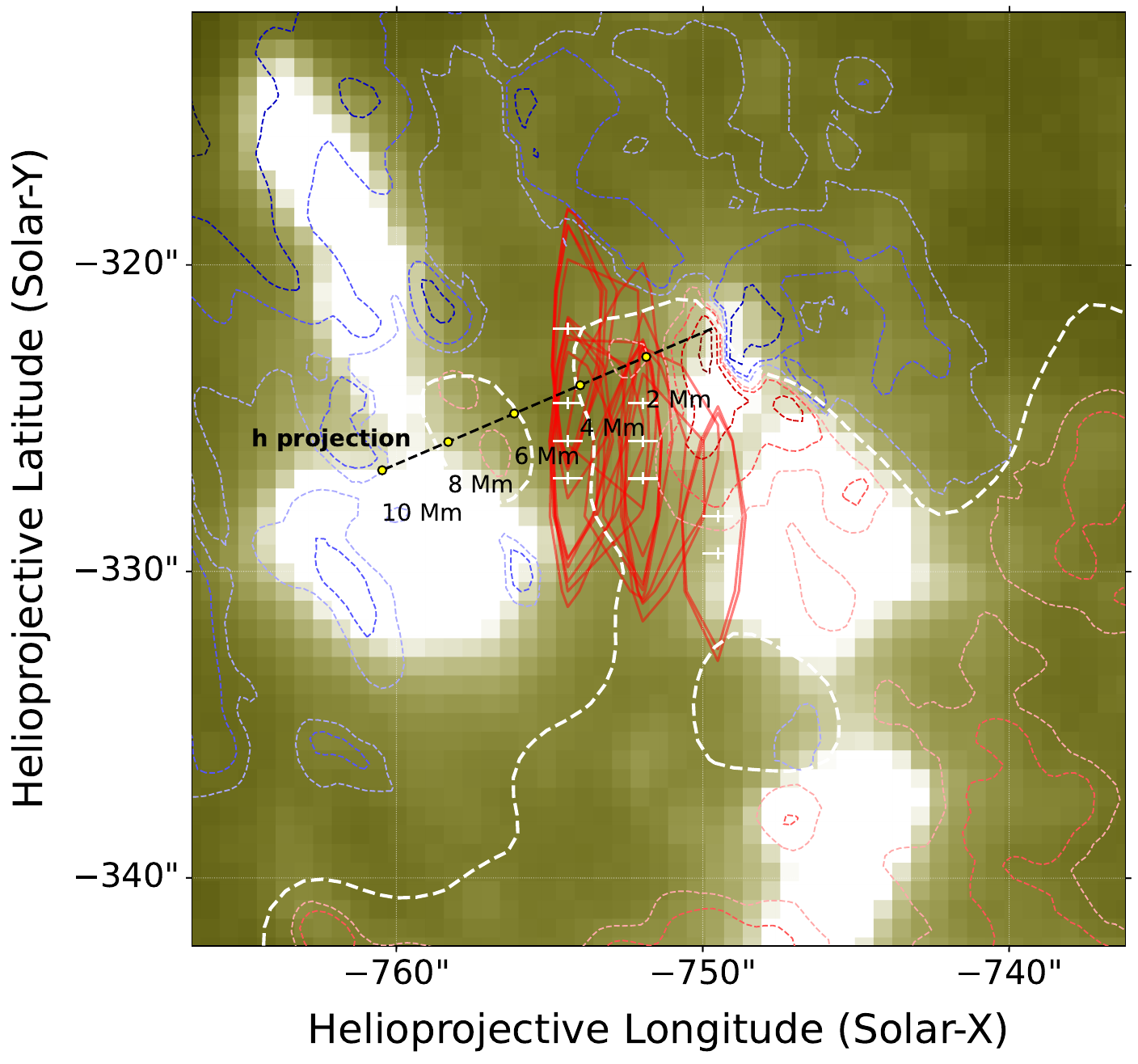}  
  \hfill
  \includegraphics[width=0.525\textwidth]{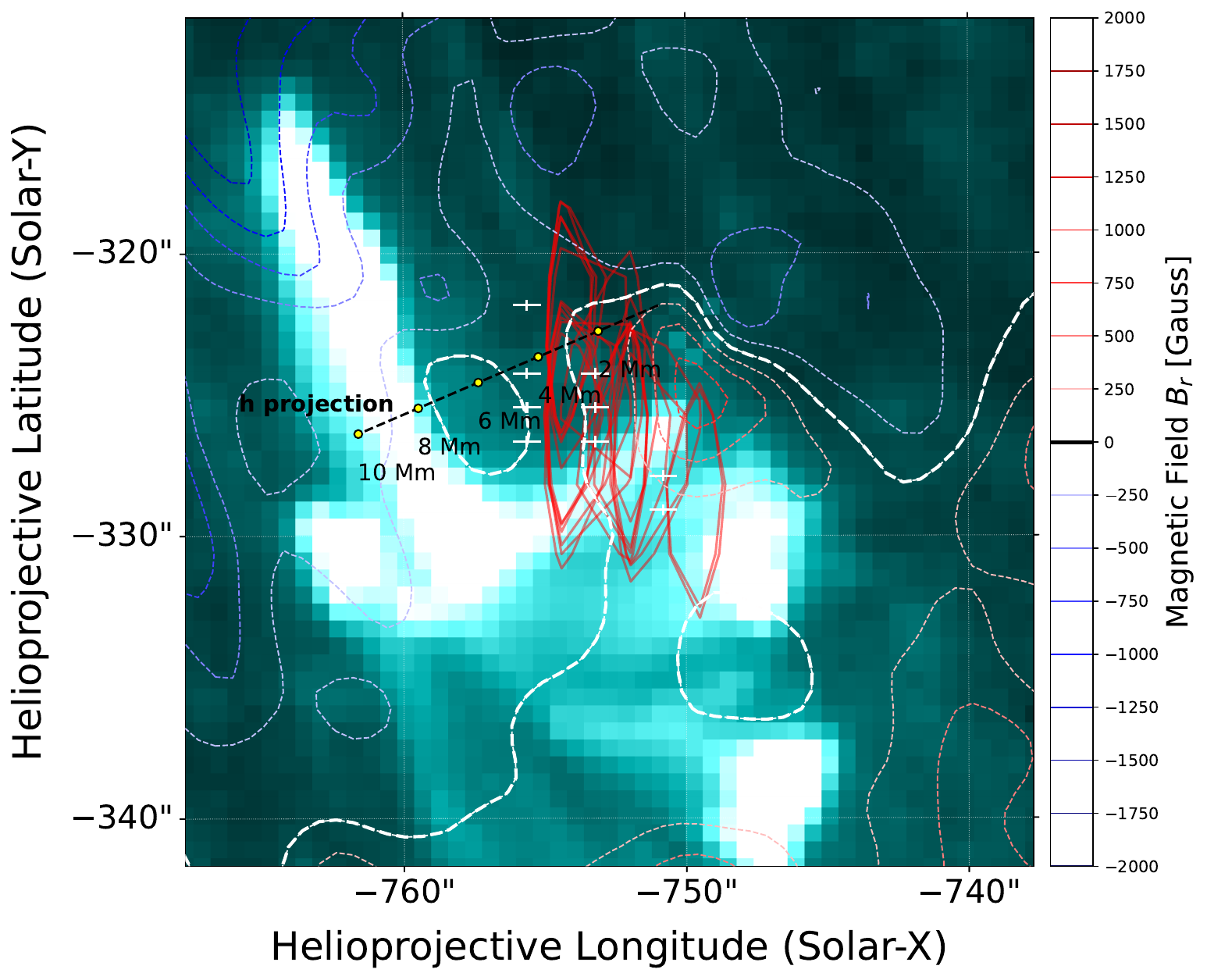}  
  \caption{Top: Apparent displacement speed of the compact MW source at 11.4\,GHz (cyan symbols, left axis) together with the peak SRH Stokes~$I$ brightness temperature (grey curve, right axis). Error bars represent the positional and temporal uncertainties. The cyan dashed line marks the flare maximum, crimson dashed lines indicate the interval of the narrowband coherent burst, and black dashed lines separate the successive impulsive episodes identified from the SXR time profile.\protect\\
  Bottom: Evolution of the compact MW source during the impulsive phase overlaid on SDO/AIA 1600\,\AA{}~(left) and 131\,\AA{}~(right) images. Filled coloured contours show the brightest (98\% of the peak intensity) background-subtracted SRH emission at successive observing times. Dashed contours denote the radial photospheric magnetic field, the thick dashed curve marks the PIL (white color), and the white crosses indicate the estimated positional uncertainty of the MW source centroid. The black dashed line indicates the projected height scale (Mm).}
  \label{MW_11_4_Speeds}
\end{figure*}

Figure~\ref{MW_11_4_Speeds} presents the temporal evolution of the apparent displacement speed of the MW source centroid at 11.4\,GHz together with the peak Stokes~$I$ brightness temperature. The source position was determined independently for each SRH image, and the apparent speed was calculated from the displacement of the source centroid between successive images divided by the corresponding time interval. Vertical error bars represent the uncertainties of the centroid positions propagated into the velocity estimates, while horizontal error bars indicate the temporal resolution of the observations. The cyan dashed line marks the time of the MW intensity maximum (02:14:53\,UTC).

The apparent source speed exhibits pronounced temporal variations throughout the impulsive phase of the flare. The largest velocities are observed during the rapid rise and around the maximum of the MW emission, indicating that the location of the most intense MW emission evolved continuously rather than remaining spatially stationary. Following the flare maximum, both the apparent source speed and the MW brightness decrease simultaneously.

The interval of the narrowband coherent burst is indicated by the crimson dashed lines. No distinct increase in the apparent source speed is observed during this episode, suggesting that the coherent burst was not associated with rapid large-scale displacement of the MW source. Instead, it is more likely related to changes in the local plasma conditions or in the electron acceleration process within an already existing compact magnetic structure.

The characteristic variations in the apparent source speed occur on timescales comparable to those of the MW quasi-periodic pulsations. Together with the systematic displacement of the source centroid (Fig.~\ref{AIA_SRH_timeevo}), this behaviour supports the interpretation that the flare energy release proceeded through a sequence of successive reconnection episodes in closely spaced magnetic structures rather than within a single stationary acceleration site.

As can be seen from the figure, the most pronounced variations in the apparent source velocity occur in the vicinity of the flare maximum. During this interval, the projected velocity of the MW source centroid reaches values in the range of approximately 200--800\,km$\cdot$\,s$^{-1}$, with an estimated uncertainty of $\pm174$\,km$\cdot$\,s$^{-1}$. Before and after the flare peak, the inferred velocities are generally lower and remain consistent with a comparatively stable source position within the measurement uncertainties.

The observed increase in the apparent source velocity near the flare maximum suggests that the spatial distribution of accelerated electrons changed most rapidly during the period of strongest energy release. Since the source size remained close to the instrumental beam width throughout the event, the detected motion is more likely associated with a displacement of the dominant MW-emitting region within a compact magnetic structure than with a physical expansion of the source itself.

\section{CORONAL MAGNETIC FIELD STRUCTURE FROM NONLINEAR FORCE-FREE FIELD EXTRAPOLATION}
The three-dimensional coronal magnetic field structure was reconstructed within the framework of a NLFFF approximation using the optimisation method developed by~\cite{Wheatland2000} and implemented in the GX Simulator software package~\cite[]{Nita2015}. As the lower boundary condition, we used an SDO/HMI vector magnetogram acquired at 01:58:36\,UTC (the midpoint of the 12-minute integration interval), prior to the onset of the investigated flare. The computational domain consisted of a grid of $160\times150\times100$ cells with a spatial resolution of 1.2\,Mm per cell, corresponding to approximately 3.3 times the linear pixel size of the original SDO/HMI data.

\begin{figure*}
\centering
\includegraphics[width=0.65\textwidth]{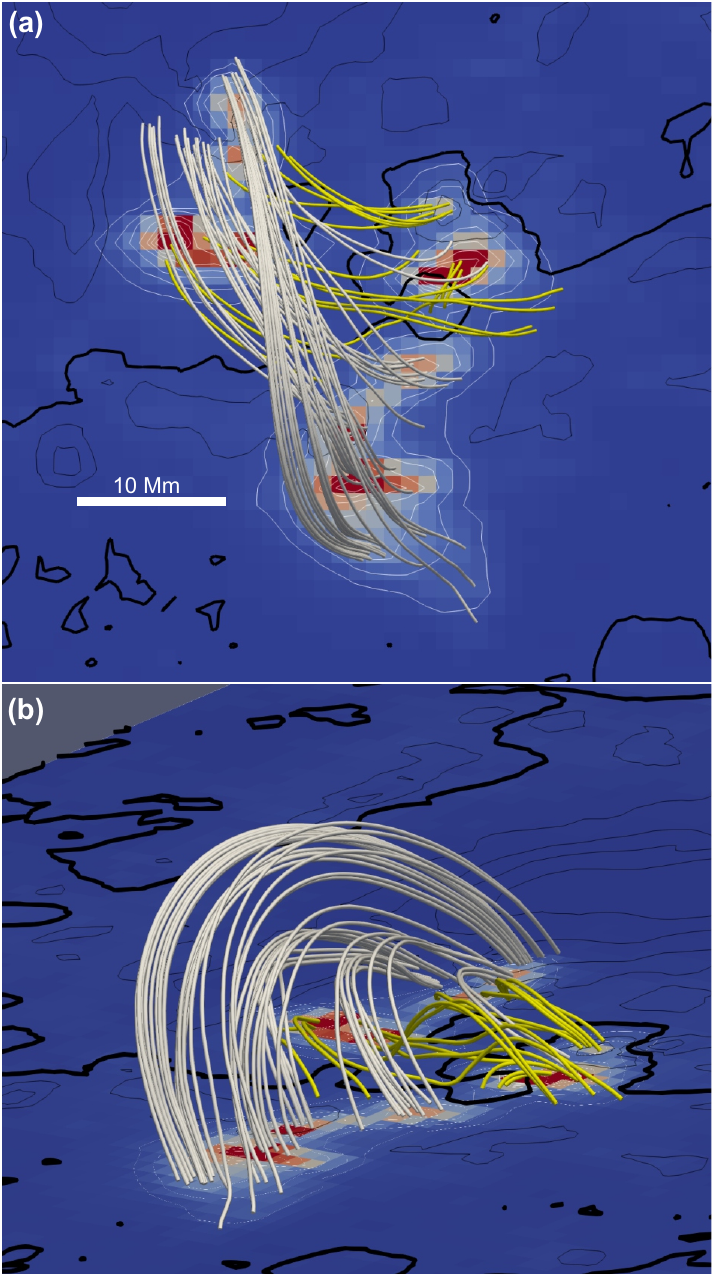}
\caption{Results of the NLFFF extrapolation of the coronal magnetic field shown from the top (a) and side (b) perspectives. Magnetic field lines were traced from the vicinity of the UV brightenings observed in the SDO/AIA\,1600\,\AA{} channel, which are displayed as the background image and outlined by thin grey contours. Yellow field lines correspond to the lowest magnetic loops, while white field lines represent higher coronal structures. The thick black curve marks the PIL. Thin black contours indicate levels of the radial magnetic field strength of $\pm1000$, $\pm1500$, and $\pm2000$\,G.}
\label{NLFFF}
\end{figure*}

The results of the NLFFF magnetic-field extrapolation are visualized in Fig.\,\ref{NLFFF} using \textsc{ParaView} from two different viewing angles. The magnetic field lines are shown overlaid on the difference UV image obtained in the SDO/AIA~1600\,\AA{} channel during the impulsive phase of the flare, with white contours indicating UV intensity isophotes. The top view (a) illustrates the projection of the magnetic field lines onto the photosphere relative to the PIL and the contours of the radial magnetic field component; a 10\,Mm scale bar is included for reference. The side view (b) provides a clearer representation of the height distribution and three-dimensional orientation of the magnetic field lines.

The overall magnetic topology shown in Fig.\,\ref{NLFFF} demonstrates good correspondence between the footpoints of the extrapolated magnetic loops and the locations of the chromospheric brightenings observed in the AIA\,1600\,\AA{} channel. Two colours are used to distinguish different magnetic structures: yellow field lines represent a system of low-lying loops, whereas white field lines correspond to longer and higher coronal loops (the separation is somewhat arbitrary). The brightest and most prominent UV brightenings located near the PIL are connected by low-lying magnetic field lines. In contrast, brightenings situated farther from the PIL are associated with higher magnetic structures. Notably, the higher the magnetic loop, the smaller its shear relative to the PIL. The lowest and most strongly sheared loops are concentrated in regions of enhanced electric current density.

The PIL exhibits a complex geometry in the vicinity of the north-western intense UV brightening, where compact magnetic loops and the strongest gradient of the radial magnetic field component $|\nabla_h B_r|$, are also located. It is noteworthy that pre-flare brightenings were observed in this region in the AIA UV images. These signatures may indicate either the emergence of a compact system of magnetic loops or shear motions of small pores and magnetic elements near the main sunspot. Such activity could have acted as the trigger of the investigated flare and appears to have developed at relatively low coronal heights ($\lesssim10$\,Mm). The impulsive phase most likely originated in the vicinity of the PIL, where the low-lying magnetic loops shown in yellow are concentrated.

\begin{figure*}
\centering
\includegraphics[width=1.0\textwidth]{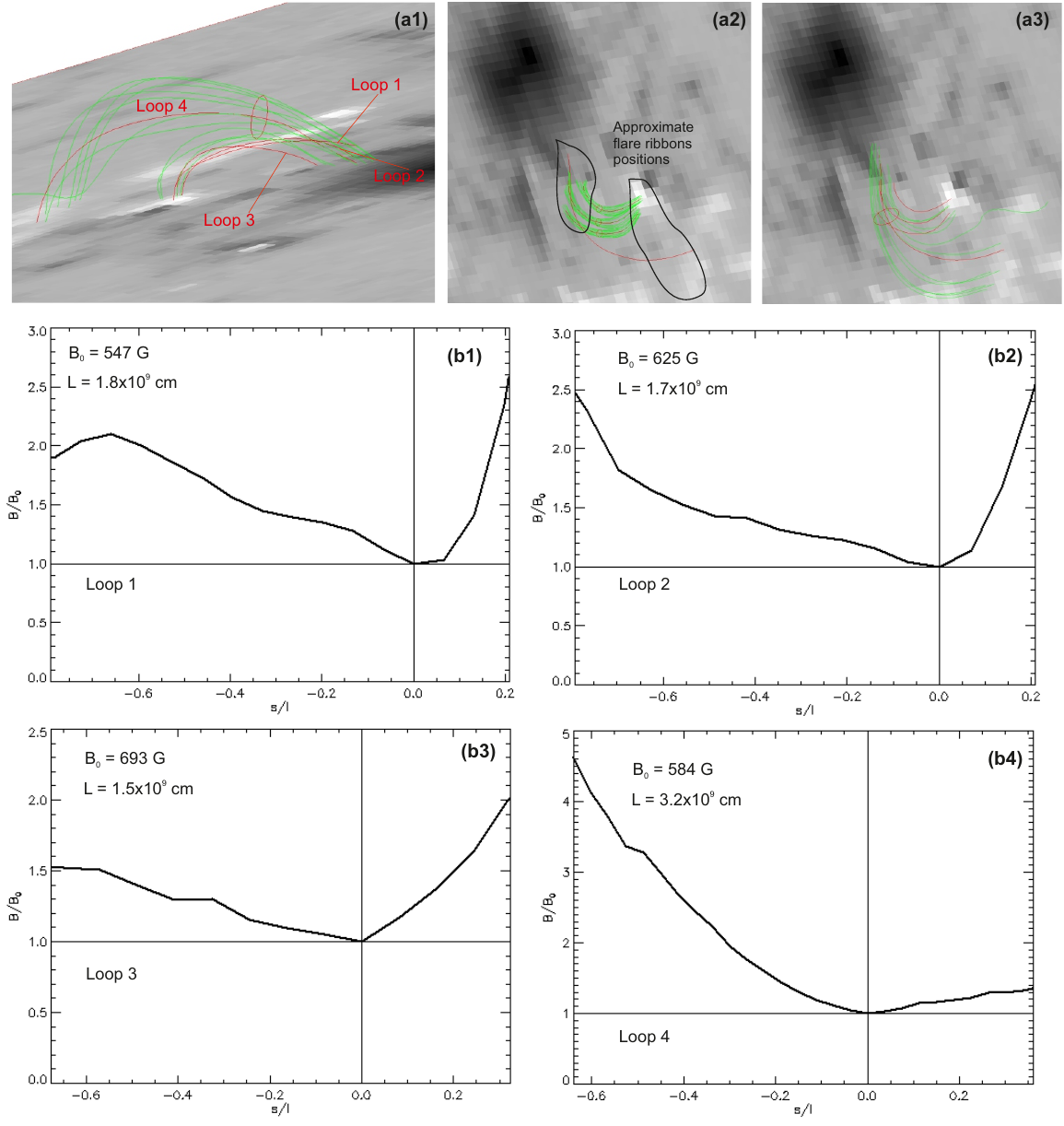}
\caption{Results of the NLFFF extrapolation visualized in GX Simulator for four representative magnetic loops with circular cross-sections. Panels (a1--a3) show the selected loops from three different viewing angles. Panels (b1--b4) present the distributions of the magnetic field strength along each loop. Horizontal and vertical lines indicate the locations and values of the magnetic-field minima (labelled within the panels). The vertical axis shows the magnetic field strength normalized to its minimum value along the loop. The horizontal axis represents the normalized distance along the loop, measured relative to the position of the magnetic-field minimum, with the total loop length normalized to unity.}
\label{loops_GX}
\end{figure*}

Figure\,\ref{loops_GX}\,(a1--a3) presents four representative magnetic loops selected manually within the flare region using the GX Simulator tools and based on the NLFFF extrapolation results. These flux tubes, represented with circular cross-sections, were chosen to illustrate both a low-lying flare loop and several relatively higher magnetic structures. The corresponding distributions of the magnetic field strength along each loop are shown in Fig.\,\ref{loops_GX}\,(b1--b4). The minimum magnetic field strengths lie within the range $B_0=547$--$693$\,G. As expected, the longer and higher loops possess weaker magnetic fields. The maximum magnetic field strengths are found near the sunspot regions of the AR. These results will be used in the following sections for physical estimates, interpretation of the MW emission spectrum, and discussion of possible mechanisms responsible for the observed QPPs.

\section{SPECTRAL ANALYSIS OF X-RAY AND MICROWAVE EMISSION}
\subsection{Microwave and X-ray Spectra Near the Flare Peak}
The impulsive HXR emission above 20\,keV and the MW emission throughout the entire observed frequency range clearly indicate the presence of a non-thermal electron population. Before discussing the temporal evolution of the thermal and non-thermal plasma parameters derived from spectral fitting, we first examine the X-ray and MW spectra obtained near the peak of the impulsive emission.

\begin{figure*}
\centering
\includegraphics[width=1.0\textwidth]{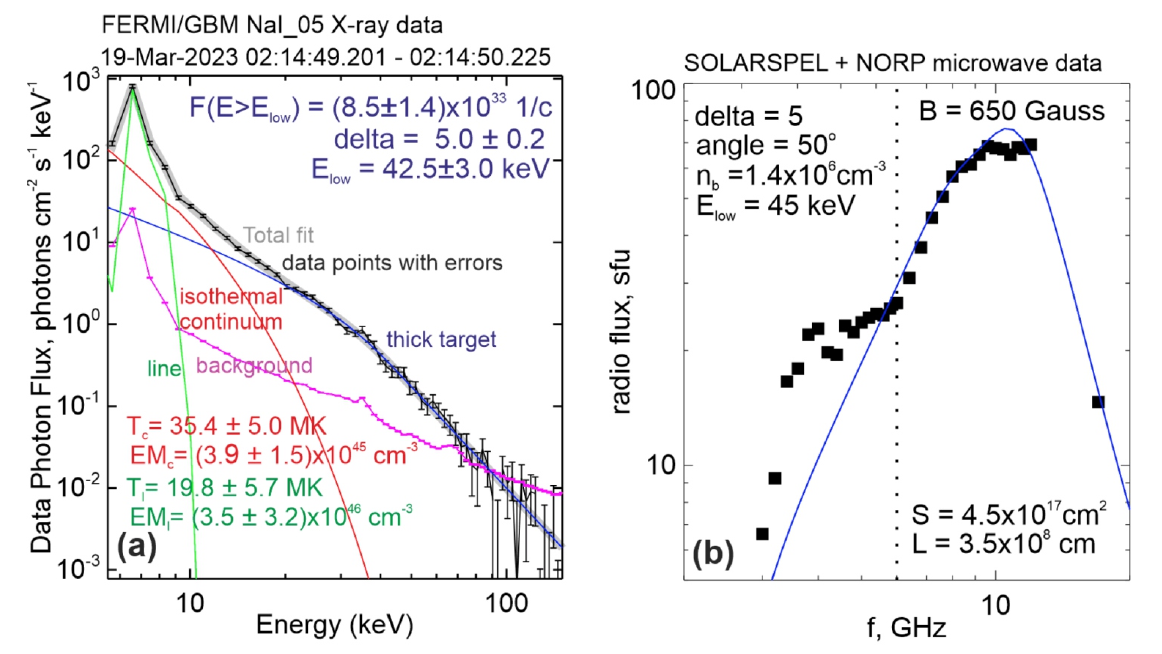}
\caption{Examples of spectral fitting near the flare maximum. (a) The X-ray photon spectrum measured by the Fermi/GBM NaI-05 detector together with the best-fit thermal and non-thermal model components. (b) The combined MW spectrum constructed from SOLARSPEL observations in the 3--12\,GHz range and NoRP observations at 17\,GHz, together with the corresponding gyrosynchrotron model fit above 6\,GHz. The X-ray panel (a) includes all fitted model components and the derived parameters with their uncertainties.}
\label{Xray_MW_peak_spec}
\end{figure*}

For the present event, the NaI-05 detector of \textit{Fermi}/GBM provides the highest-quality X-ray spectra up to $\sim100$\,keV in terms of the signal-to-noise ratio. Count-rate spectra are available with a temporal resolution of 1--4\,s. An example of the background-subtracted photon spectrum over the energy range 6--120\,keV, accumulated during a 1\,s interval near the HXR peak, is shown in Fig.,\ref{Xray_MW_peak_spec}a. The spectrum is fitted with a three-component model consisting of:
\begin{enumerate}
\item A single-temperature thermal bremsstrahlung continuum characterised by the plasma temperature ($T_c$) and emission measure ($EM_c$). Throughout this work, these parameters are used to characterise the hot plasma directly associated with the primary flare energy-release region and the initial stage of electron acceleration.
\item A single-temperature thermal line-emission component representing the Fe/Ni line complex, with free parameters $T_l$ and $EM_l$. This component is introduced to improve the fit below $\sim10$\,keV, thereby preventing systematic deviations at low energies from affecting the inferred non-thermal component. Owing to the relatively large uncertainties of $T_l$ and $EM_l$, these parameters are not used in the subsequent physical interpretation. The corresponding emission most likely originates either from cooler plasma ($T\approx10$--20\,MK) produced by secondary hydrodynamic processes or from cooler components of the multi-thermal flare plasma. In contrast, the hot plasma component described by ($T_c$, $EM_c$) is determined more reliably and is therefore considered of greater physical significance. We suggest that it is directly related to magnetic reconnection and electron acceleration processes.
\item A non-thermal bremsstrahlung component produced by accelerated electrons within the thick-target approximation. The free parameters are the total electron flux integrated above the low-energy cutoff, $F(E>E_{\mathrm{low}})$, the electron spectral index ($\delta$), and the low-energy cutoff energy ($E_{\mathrm{low}}$). The electron spectrum is assumed to extend as a single power law to arbitrarily high energies. The thick-target model is motivated, in particular, by the presence of paired HXR sources located at opposite footpoints of the flare loop in which the impulsive-phase energy release occurred.
\end{enumerate}

The adopted X-ray spectral model is not unique. Alternative parameterizations can provide comparably good statistical fits. However, such models generally require a larger number of free parameters, leading to increased model complexity, stronger parameter degeneracy, and less stable temporal evolution of the fitted parameters, including artificial discontinuities caused by transitions between different local minima during the optimisation procedure. We therefore adopt the above three-component model as the baseline representation of the X-ray spectra, since it provides the simplest physically motivated description of the dominant processes in the flare while ensuring robust and self-consistent parameter estimates throughout the analysed time interval.

Figure~\ref{Xray_MW_peak_spec}a also shows the best-fit parameters of the thermal plasma and accelerated-electron spectrum, together with their uncertainties, as obtained using the OSPEX software package (IDL). An important feature of the fitted spectrum is the well-defined break at approximately 40\,keV, which is naturally interpreted as the low-energy cutoff of the accelerated-electron distribution \citep[e.g.][]{Sui2007}. In many solar flares, this cutoff is effectively masked by the broad thermal component of the X-ray spectrum and therefore cannot be reliably constrained. Consequently, it is often fixed to an assumed value during spectral fitting \citep[e.g.][]{Sharykin2015}. In the present event, however, the low-energy cutoff is clearly resolved and can be treated as a free fitting parameter with only moderate uncertainties. This provides a more reliable basis for comparing the properties of the thermal plasma with those of the accelerated-electron population.

The combined MW spectrum covering the frequency range 3--17\,GHz (SOLARSPEL~3--12\,GHz; NoRP~17\,GHz) is shown in Fig.\,\ref{Xray_MW_peak_spec}b. The observed spectrum clearly consists of two components separated at approximately $\approx6$\,GHz. In the present work, we model only the high-frequency component because it is well reproduced by a homogeneous gyrosynchrotron source. This choice is also supported by the SRH images at frequencies above 6\,GHz, which consistently show a single compact source located between the flare ribbons and most likely associated with the coronal sections of the flare loops. The low-frequency component is likely produced by a different population of emitting electrons associated with larger magnetic loops and substantially weaker magnetic fields. Modelling this component would therefore require an additional source with different physical parameters. Since the NLFFF extrapolation yields magnetic-field strengths of at least $\sim600$\,G in the compact flare loops (Fig.,\ref{loops_GX}), we restrict our analysis to the high-frequency gyrosynchrotron emission, which can be adequately described by a single homogeneous source. The model spectra are calculated using the fast gyrosynchrotron codes developed by \cite{FleishmanKuznetsov2010}.

In contrast to the X-ray domain, the gyrosynchrotron MW spectrum of a homogeneous source depends explicitly on the magnetic-field strength and orientation, as well as on the source geometry (projected area and line-of-sight depth). The parameters adopted for the spectral fitting are listed in Fig.\,\ref{Xray_MW_peak_spec} and include the electron spectral index, the angle between the magnetic field and the line-of-sight, the magnetic-field strength, the non-thermal electron density, the low-energy cutoff, and the source geometry. The inferred magnetic-field strength is consistent with the field expected near the loop-top of the lower loops (Fig.\,\ref{loops_GX}b2--b3). Since these loops exhibit a magnetic mirror ratio of $B_{max}/B_{min}\sim2$--3, a fraction of accelerated electrons is expected to remain trapped in the coronal sections of the loops where the magnetic field is close to its minimum.

An upper estimate of the trapping time, $\tau_{\rm trap}$, for electrons with energies of 42.5\,keV (the fitted value of $E_{\rm low}$) and 100\,keV can be obtained from Eq.~12.5.11 of~\cite{book_Aschwanden2005},
\begin{equation}
\label{t_trap} 
\tau_{trap} \lesssim 0.95\times10^8\frac{E_{keV}^{3/2}}{n_e}\frac{20}{ln\Lambda}\nonumber
\end{equation}
which is derived in the weak-diffusion approximation for Coulomb collisions. Here Coulomb logarithm $\ln\Lambda=22.7$ for $T_c=35.4$\,MK, while the thermal electron density is estimated as $n_e=\sqrt{EM_c/V}\approx4.9\times10^9$\,cm$^{-3}$ using a source volume of $V\approx1.6\times10^{26}$\,cm$^3$, obtained by multiplying the approximate loop length ($1.6\times10^9$\,cm) by the cross-sectional area ($\sim10^{17}$\,cm$^2$) of the brightest AIA 1600\,\AA{} ribbon kernel. These estimates yield $\tau_{\rm trap}\lesssim5$--18\,s, comparable to both the overall duration of the MW burst and the characteristic duration of its individual peaks. Since gyrosynchrotron emission is produced most efficiently by electrons with energies exceeding $\sim100$\,keV, the observed MW time profile is therefore consistent with collisional trapping in the flare loops. Even longer trapping times may be expected for the highest-energy electrons, whereas plasma turbulence could shorten the effective trapping time through wave-particle scattering.

The observed MW spectrum in the 6--17\,GHz range is reproduced using a homogeneous gyrosynchrotron source model \citep[]{Dulk1985, FleishmanKuznetsov2010}. The magnetic field strength was fixed at 650\,G according to the NLFFF extrapolation, while the low-energy cutoff was set to 45\,keV, consistent with the HXR thick-target fit. The power-law index of the accelerated-electron distribution was fixed at $\delta=5$, matching the value derived from the HXR spectrum. The geometrical dimensions of the model loop were chosen based on several observational considerations. First, the MW spectrum is reasonably well reproduced by a homogeneous source model, suggesting that the spatial variation of the emitting electron population is not excessively large within the flare region. This allows the source to be approximated by a single magnetic loop, for example, assuming that the non-thermal electrons are concentrated near its central section. To estimate the projected length of the MW source, we adopted the length of Loop~3 shown in Fig.\,\ref{loops_GX}\,b3 and slightly reduced it to account for projection effects, obtaining $l\approx1.3\times10^9$\,cm (see the spatial scale in Fig.\,\ref{NLFFF}). The loop cross-section was estimated from the characteristic sizes of the two UV flare kernels associated with the low-lying loops (shown in yellow in Fig.\,\ref{NLFFF}) observed in the AIA\,1600\,\AA{} channel, yielding $L\approx3.5$,Mm. The resulting projected source area is therefore approximately $S \approx lL \approx 4.5\times10^{17}\ \mathrm{cm}^{2}$.

The relationship between the integrated electron flux above the low-energy cutoff, $F(E>E_{\mathrm{low}})$, and the density of accelerated electrons is given by
\begin{equation}
\label{FE_con} 
F(E>E_{low}) = 2n_bS\cdot\frac{\delta - 3/2}{\delta - 2}\cdot\sqrt{E_{low}\frac{2}{m_e}},\nonumber
\end{equation}
where $n_b$ is the number density of non-thermal electrons derived from the gyrosynchrotron spectral fit, $m_e$ is the electron mass, and $S$ is the cross-sectional area of the magnetic loop through which accelerated electrons precipitate into the lower layers of the solar atmosphere. The factor of two accounts for precipitation into both footpoints of the loop. Using the parameters obtained from the MW spectral modelling, we obtain $F(E>E_{low}) \approx 5\times 10^{33}$\,electrons$\cdot$s$^{-1}$, which is consistent, to within the uncertainties of the source geometry, with the value derived from the HXR thick-target fit, $F(E>E_{low}) \approx 8.5\times 10^{33}$\,electrons$\cdot$\,s$^{-1}$. Thus, the parameters of the accelerated electron population inferred from the MW and HXR spectra are mutually consistent. The remaining discrepancy can plausibly be attributed to uncertainties in the source geometry and to transport effects that are not included in the homogeneous-source model, such as magnetic trapping and pitch-angle scattering.

\subsection{Temporal Evolution of Thermal-Plasma and Non-Thermal-Electron Parameters}
\begin{figure*}
\centering
\includegraphics[width=0.65\textwidth]{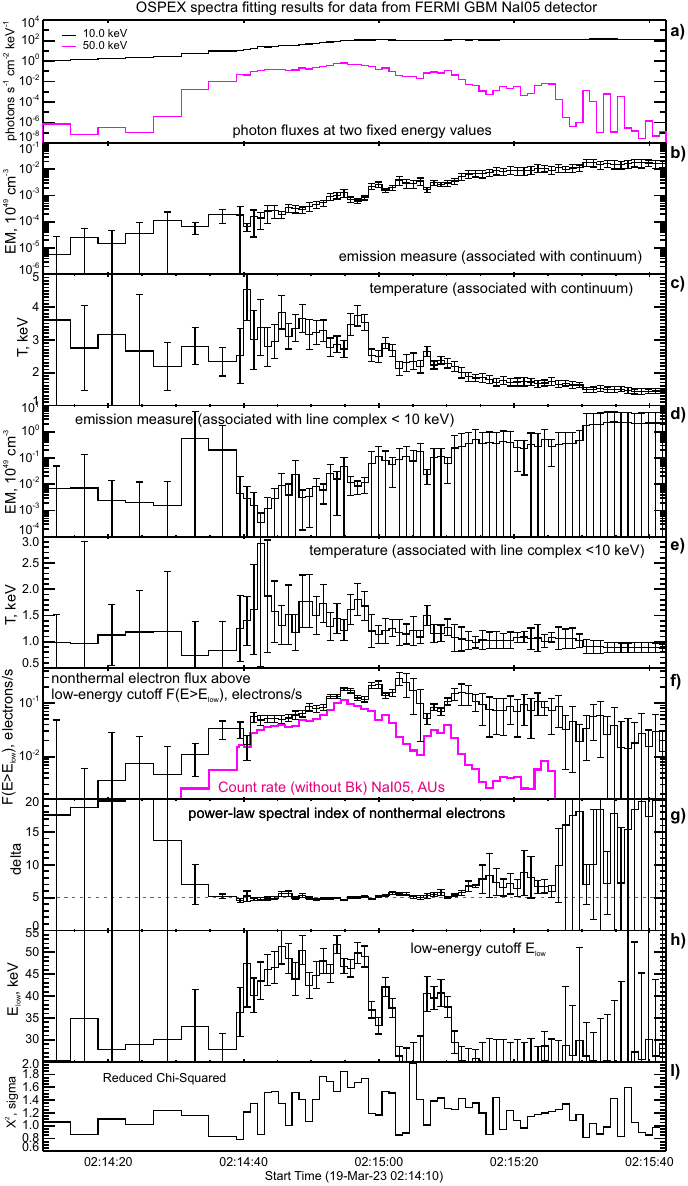}
\caption{Results of the spectral fitting of the Fermi/GBM NaI-05 X-ray observations with a temporal resolution of 1--4\,s. (a) The model photon fluxes at 10 and 50\,keV. (b,c) The parameters of the single-temperature thermal bremsstrahlung component. (d,e) The parameters of the single-temperature line-emission component associated with the Fe/Ni complex. (f--h) The parameters of the non-thermal thick-target model describing accelerated electrons with a power-law energy distribution. (i) The goodness-of-fit parameter, $\chi^{2}$.}
\label{Xray_Fit_re}
\end{figure*}

The three-component fitting approach applied to the Fermi/GBM (NaI-05 detector) X-ray spectra was introduced in the previous subsection. Figure~\ref{Xray_Fit_re} presents the results of spectral fitting for the entire time series in the 6--120\,keV energy range. Owing to the high count rates during the flare, spectra of sufficient quality were obtained with temporal resolutions as short as 1\,s. Prior to approximately 02:14:40\,UTC, the temporal cadence of the fitted spectra was 4\,s. The figure shows the temporal evolution of all principal free parameters used in the fitting procedure. The overall quality and stability of the spectral fits can be assessed from the reduced $\chi^{2}$ values shown in Fig.\,\ref{Xray_Fit_re}\,l.

The spectral analysis revealed several important characteristics of electron acceleration and plasma heating during the impulsive phase of the flare:
\begin{enumerate}
\item The spectral index of the accelerated-electron energy distribution remained relatively stable throughout the impulsive phase, with $\delta \approx 5.0 \pm 0.4$.
\item The low-energy cutoff of the accelerated-electron spectrum was found to vary within the range $E_{\mathrm{low}}=$ 30--52\,keV, with a characteristic average value of approximately 45\,keV.
\item Distinct pulses were detected in both the plasma temperature and the integrated flux of accelerated electrons, with characteristic durations of about 5\,s.
\item The thermal plasma temperature varied between approximately 25 and 40\,MK.
\item During the rising phase of the impulsive emission, the emission measure and the flux of accelerated electrons exhibited similar temporal evolution. At later times, however, the emission measure continued to increase while the non-thermal electron flux decreased.
\item A similar temporal behaviour was also observed for the plasma temperature and the low-energy cutoff during the rise phase of the impulsive flare emission.
\end{enumerate}

\begin{figure*}
\centering
\includegraphics[width=0.8\textwidth]{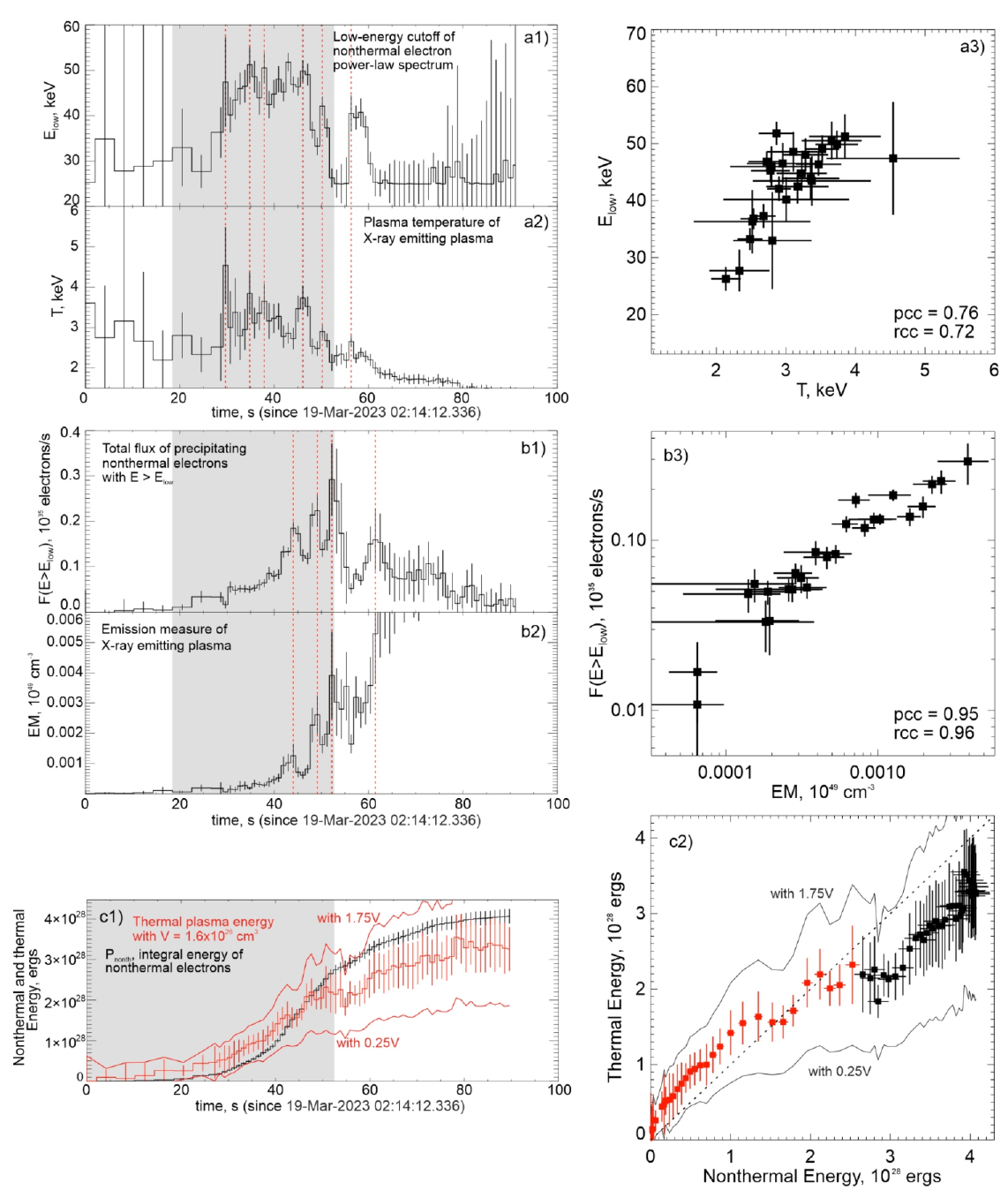}
\caption{Comparison of the parameters of the thermal plasma and the accelerated-electron population. Panels (a1--a3) show the relationship between the low-energy cutoff, $E_{\mathrm{low}}$, and the plasma temperature, $T$. Panels (b1--b3) compare the total flux of accelerated electrons with the thermal emission measure, $EM$. Panel (c1) shows the temporal evolution of the cumulative non-thermal electron energy (black) and the thermal energy of the plasma (red). The thin solid red curves in panels (c1,c2) indicate the thermal energy calculated for plasma volumes decreased and increased by 75\%, respectively. Panels (a3), (b3), and (c2) show the correlations between the corresponding parameters within the time interval highlighted by the grey shading in panels (a1,a2), (b1,b2), and (c1). In panel (c2), the dashed line denotes the one-to-one relation. Red symbols correspond to data points within the grey-shaded interval, while black symbols represent measurements outside this time range.}
\label{Xray_Energy}
\end{figure*}

During the early stage of the impulsive phase, a nearly simultaneous increase in the logarithm of the thermal-plasma emission measure (normalized to $10^{49}$\,cm$^{-3}$) and the logarithm of the total flux of accelerated electrons (normalized to $10^{35}$\,electrons\,s$^{-1}$) is observed. Their relationship can be approximated by
\begin{equation}
\label{ME_low}
lgF_{35} = (0.69\pm0.05)lgEM_{49}+(1.18\pm0.15)\nonumber
\end{equation}
or, in simplified form,
\begin{equation}
\label{ME_rub}
F_{tot}(E>E_{low})_{35}\sim 10\sqrt{EM_{49}}\nonumber
\end{equation}
Moreover, the fine temporal structures of both quantities closely follow each other up to the peak of the accelerated-electron flux (the corresponding time interval is indicated by the grey shading in Fig.\,\ref{Xray_Energy}). The correlation between these parameters is illustrated in Fig.\,\,\ref{Xray_Energy}b3, which includes only the data points within the shaded interval. Restricting the analysis to the earliest stage of the impulsive phase, corresponding to $EM > 9\times10^{45}$\,cm$^{-3}$ and including the first burst of accelerated electrons, yields the following relationship based on 19 data points:
\begin{equation}
\label{ME_19}
lgF_{35} = (0.86\pm0.10)lgEM_{49}+(1.82\pm0.35)\nonumber
\end{equation}
which can be approximately expressed as
\begin{equation}
\label{ME_49}
F_{0}(E>E_{low})_{35}\sim F_{tot35}^2 \sim 100EM_{49}\nonumber
\end{equation}
Here, $F_{\mathrm{tot},35}$ refers to the entire interval from the beginning of the impulsive phase up to the peak electron flux, whereas $F_{0}$ corresponds to a shorter interval encompassing only the initial rise and the first burst of accelerated electrons (up to approximately 45\,s in Fig.\,\ref{Xray_Energy}).

The latter relationship indicates that the increase in the accelerated-electron flux was initially more rapid than the increase in the emission measure. This behaviour may suggest that the earliest stage of the impulsive phase was dominated by direct plasma heating and particle acceleration, while chromospheric evaporation and other hydrodynamic responses had not yet fully developed. At later times, the continued increase of the emission measure, accompanied by a decrease in the accelerated-electron flux, is consistent with progressive filling of the flare loop by evaporated chromospheric plasma. Such an increase in the ambient plasma density may reduce the efficiency of the acceleration process and/or modify the transport properties of the accelerated electrons.

Taking into account the normalization factors of $10^{35}\,\mathrm{electrons ^{-1}}$ and $10^{49}$\,cm$^{-3}$, the latter relationship can be rewritten as
\begin{equation}
\label{equal_1}
F_{0}(E>E_{low})\sim {10^{-12}\,\mathrm{[cm^3 \cdot s^{-1}]}} \cdot EM\nonumber
\end{equation}
The coefficient $10^{-12}\,\mathrm{cm^{3} \cdot s^{-1}}$ has the dimensions of volume multiplied by frequency. It may therefore be interpreted as a characteristic measure of the efficiency with which electrons from the hot parent plasma are transferred into the accelerated population. A more detailed discussion of the possible physical meaning of this coefficient is presented in chapter~\textit{Discussion}.

The early impulsive phase also exhibits a relationship between the plasma temperature and the low-energy cutoff of the accelerated-electron spectrum:
\begin{equation}
\label{ME_Elow}
E_{low} = (9.50\pm1.90)T+(13.7\pm5.9)\nonumber
\end{equation}
where $E_{\mathrm{low}}$ is measured in keV and $T$ in MK. For illustrative purposes, this dependence may be approximated by
\begin{equation}
\label{ME_Elow2}
E_{low}-10 \text{\,keV}\sim10T \text{\,keV}\nonumber
\end{equation}
The correlation between these parameters is shown in Fig.\,\ref{Xray_Energy}\,a3. As in the previous panel, only the data points within the grey-shaded interval, extending from the onset of the impulsive phase to the peak of the accelerated-electron flux, are included. The correlation is weaker than that found for the $lgF$--$lgEM$ relationship, with a Pearson correlation coefficient of 0.76. The observed dependence between $E_{\mathrm{low}}$ and $T$ suggests a close coupling between plasma heating and the formation of the non-thermal electron population during the early impulsive phase. In particular, the low-energy cutoff appears to scale approximately with the characteristic thermal energy of the hot plasma. This behaviour is consistent with scenarios in which the accelerated electrons originate from the high-energy tail of the evolving thermal distribution, although alternative interpretations involving a common underlying acceleration process cannot be excluded.

Figures\,\ref{Xray_Energy}c1 and c2 show the temporal evolution of the cumulative energy of accelerated electrons, $E_{\mathrm{nonth}}$, and the thermal energy of the X-ray-emitting plasma, $U_{\mathrm{th}}$, respectively. The thermal energy was estimated as
\begin{equation}
\label{U_th}
U_{th}(t)= 3k_BT(t)\sqrt{EM(t)\cdot V}\nonumber
\end{equation}
where $k_B$ is the Boltzmann constant and the source volume was assumed to be constant, $V \approx 1.6\times10^{26}$\,cm$^{3}$, with an arbitrary uncertainty of approximately $\pm75\%$. Given the confined nature of the flare and the absence of detectable flare-ribbon motion, the assumption of a nearly constant source volume appears reasonable as a first approximation. The adopted volume\,$V$ was derived from the geometric parameters used in the gyrosynchrotron spectral modelling and from the assumption that, during the early impulsive phase, the brightest X-ray source was associated with a low-lying magnetic loop rooted in the most intense chromospheric brightenings near the PIL (Figs.\,\ref{NLFFF} and \ref{loops_GX}).

The cumulative energy deposited by non-thermal electrons was calculated as
\begin{equation}
\label{Eng_nonthE}
E_{nonth}(t) = \int_{t_0}^{t}F(t)E_{low}(t)\frac{\delta(t)-1}{\delta(t)-2}dt\nonumber
\end{equation}

For the adopted volume of $V = 1.6\times10^{26}$\,cm$^{3}$, the accumulated non-thermal energy remains below the thermal energy up to approximately 45\,s after the onset of the impulsive phase. At later times, $E_{nonth}$ exceeds $U_{th}$. However, given the substantial uncertainty in the source volume, a more robust conclusion is that the thermal and non-thermal energy contents are of the same order of magnitude throughout the impulsive phase, i.e. $E_{nonth}\approx U_{th}$.
The correlations identified between the properties of the accelerated-electron population and those of the thermal plasma during the earliest stage of the flare suggest a close coupling between plasma heating and particle acceleration before large-scale flare hydrodynamics become fully established. The characteristic timescales of chromospheric evaporation and plasma transport are expected to exceed those of the initial energy-release episodes. Consequently, the observed relationships may provide information on the primary acceleration process operating in the hot flare plasma. The physical implications of these findings are discussed in the following section.

\section{DISCUSSION}
\label{sec:discussion}
\subsection{Event morphology and applicability of the standard flare model}
The spatial distribution of the emission sources together with the reconstructed magnetic-field configuration provides an opportunity to assess the extent to which the standard CSHKP flare model describes the present event. The flare exhibits several characteristics of a compact confined flare. Although two chromospheric flare ribbons and a hot coronal loop system are observed, the overall morphology differs from that expected for a classical eruptive two-ribbon flare.

The NLFFF extrapolation reveals a compact system of low-lying magnetic loops, with heights below approximately 10\,Mm, concentrated in the vicinity of the PIL. The flare ribbons identified in the SDO/AIA observations do not show the systematic separation from the PIL that is characteristic of the standard CSHKP scenario involving the progressive formation of increasingly higher post-flare loops. Instead, the brightest emission remains confined to a relatively small region throughout the impulsive phase.

The reconstructed magnetic topology and the morphology of the emission sources suggest that the flare was most likely driven by magnetic reconnection within a compact system of interacting low-lying loops located near the PIL. Such a configuration is more naturally interpreted in terms of loop--loop interaction than by the classical large-scale eruptive flare scenario. The absence of any evidence for a coronal mass ejection or large-scale expansion of the flare arcade further supports this interpretation.

Nevertheless, the observed morphology does not contradict the fundamental concepts of the standard flare model. The presence of conjugate flare ribbons, hot post-reconnection loops, and compact HXR and MW sources indicates that magnetic reconnection remained the primary mechanism of energy release. The principal difference is that the reconnection appears to have been confined to a compact low-altitude magnetic system rather than occurring beneath an erupting flux rope.

Instead, the observations appear to be more consistent with energy release occurring within a compact system of interacting low-lying loops. Such a scenario is supported by the distribution of localized brightenings, the morphology of the flare ribbons, and the geometry of the magnetic field reconstructed from the NLFFF extrapolation.

A possible interpretation of the temporal and spatial evolution of the flare is as follows. Subphotospheric motions, such as localized magnetic-flux emergence or relative displacements of magnetic elements, may have increased magnetic stress in the vicinity of the PIL, particularly near the region characterized by a strong horizontal gradient of the radial magnetic field component ($\nabla_h B_r \approx 1\,\mathrm{kG \cdot Mm^{-1}}$). Under such conditions, magnetic reconnection could have been initiated between several compact low-lying magnetic loops ($h\lesssim10$\,Mm). As the magnetic configuration evolved, the energy release may have progressively involved neighbouring and higher magnetic structures characterized by smaller shear angles relative to the PIL.

In this picture, magnetic reconnection develops within current sheets formed by interacting magnetic loops whose field lines are inclined with respect to one another rather than strictly anti-parallel. The magnetic configuration appears to have remained sufficiently stable to prevent the formation and eruption of a large-scale flux rope. Consequently, the event remained confined and no CME or large-scale restructuring of the AR was observed.

The energy release occurred in a sequence of impulsive episodes with characteristic time intervals of approximately 10--15\,s, consistent with the observed QPP signatures. These episodes likely involved successive groups of magnetic field lines over a total duration of about one minute. Such behaviour naturally explains both the absence of large-scale ribbon separation and the compact morphology of the flare.

Although the available UV/EUV observations do not provide sufficient temporal resolution to directly track the evolution of individual brightenings, it is conceivable that observations with substantially higher cadence would reveal a progressive activation of magnetic structures located at increasing distances from the PIL. In this case, the flare development would correspond to a transition from the most strongly sheared low-lying loops aligned along the PIL to progressively higher and less-sheared loops spanning the PIL.

\subsection{Coupling Between Electron Acceleration and Plasma Heating}

A combined analysis of the X-ray and MW observations revealed several empirical relationships between the properties of the accelerated electrons and the hot plasma during the early impulsive phase of the flare, prior to the hard X-ray peak. The main results can be summarised as follows:
\begin{enumerate}
\item Electron acceleration proceeds in a highly non-stationary manner, consisting of successive energy-release episodes with characteristic durations of approximately 5\,s.
\item The power-law index of the accelerated-electron spectrum remains remarkably stable throughout the early impulsive phase, with $\delta \approx 5.0 \pm 0.5$.
\item The low-energy cutoff of the accelerated-electron spectrum is approximately proportional to the plasma temperature,
\begin{equation*}
E_{low} - 10 \text{\,keV} \sim 10 \times T\,\mathrm{keV}
\end{equation*}
where $T$ is expressed in MK.
\item The total flux of accelerated electrons is approximately proportional to the plasma emission measure,
\begin{equation*}
    F \sim {10^{-12}\,\mathrm{[cm^3 \cdot s^{-1}] }}\cdot EM.
\end{equation*}
\item The cumulative energy of the accelerated electrons is comparable to the thermal energy of the hot plasma, $E_{nonth} \sim U_{th}$.
\end{enumerate}

From our perspective, the relationships listed above provide important constraints on the physics of particle acceleration and energy release during the earliest stage of the flare. In particular, they indicate a close coupling between the thermal plasma and the accelerated-electron population. The relationship between $E_{\mathrm{low}}$ and $T$ may contain information about the physical conditions under which the non-thermal electron spectrum forms from the ambient hot plasma. Determining the low-energy cutoff is of particular importance because uncertainties in $E_{\mathrm{low}}$ directly translate into large uncertainties in the total energy content of accelerated electrons. Several approaches for estimating $E_{\mathrm{low}}$ have been proposed in the literature \citep[e.g.][]{Aschwanden2019}. One of the most widely used is the warm-target model \citep[]{Kontar2015}, which predicts an approximately linear relationship of the form $E_{low}\sim \delta \cdot T$. In the present study, we also find an approximately linear dependence between the low-energy cutoff and the plasma temperature, although in a somewhat different form. At present, we cannot directly relate the observed relationship to the predictions of the warm-target model. It should be noted that the warm-target approach assumes that the hard X-ray spectrum is produced in a sufficiently hot and dense target where accelerated electrons lose most of their energy through Coulomb collisions.

In contrast, our analysis focuses on the earliest stage of the impulsive phase, when large-scale flare hydrodynamics and chromospheric evaporation are likely not yet fully developed. Under such conditions, the density of the hot plasma throughout the flare volume may be insufficient for the assumptions of the warm-target model to be fully satisfied. Therefore, the observed $E_{\mathrm{low}}(T)$ relationship may reflect either the initial formation of the accelerated-electron population from the high-energy tail of the thermal distribution or a more general connection between plasma heating and electron acceleration. Distinguishing between these possibilities requires further investigation.

The relationship between the emission measure and the flux of accelerated electrons provides insight into the properties of the acceleration process operating during the early impulsive phase of the flare, while the density and emission measure of the thermal plasma evolve simultaneously. The empirical coefficient $\gamma = 10^{-12}$\,cm$^{3}$\,s$^{-1}$ obtained from
\begin{equation*}
F_{0}(E>E_{low})\sim {10^{-12}\,\mathrm{[cm^3 \cdot s^{-1}]}} \cdot EM
\end{equation*}
may characterize a characteristic acceleration rate per unit plasma volume. Using the relation
$$F\sim \gamma EM=\gamma n^2SL$$
and expressing the total electron flux as $F=fS$, where $f$ is the flux density of accelerated electrons (cm$^{-2}$\,s$^{-1}$), one obtains
$$f\sim \gamma n^2L = \gamma n N_L = \nu_{acc}N_L$$
where $N_L=nL$ is the column density of particles within the acceleration region and $\nu_{\rm acc}=\gamma n$ has the dimension of a frequency. This quantity may be interpreted as an effective acceleration probability per unit time for an electron belonging to the parent thermal population. In this interpretation, the flux density of accelerated electrons is proportional to the product of the number of available particles and their probability of being accelerated.

For a representative plasma density of $n \approx 4.9\times10^{9}$\,cm$^{-3}$, corresponding to the flare peak, the inferred acceleration rate is $\nu_{\rm acc}\approx5\times10^{-3}$\,s$^{-1}$. Over a characteristic timescale of $\tau\approx50$\,s, this corresponds to an acceleration probability of order $\nu_{\rm acc}\tau\approx0.25$. Thus, roughly one quarter of the electrons belonging to the hot parent population could participate in the acceleration process during the early impulsive phase.

At the peak of the non-thermal emission, the electron mean free path estimated from the thermal velocity, $v_{\rm Te}=3.89\times10^{5}\sqrt{T}$, and the collisional timescale, $\tau_{\rm Te}=2.75\times10^{-1}T^{3/2}/(n\Lambda)$, is
$\lambda_{mfp}= v_{Te}\tau_{\rm Te}\approx 12$\,{\rm Mm}
which is comparable to the characteristic length of the flare loop. The correlations $F(EM)$ and $E_{\rm low}(T)$, as well as the approximate energy balance $E_{\rm nonth}\sim U_{\rm th}$, were found specifically during this stage of the flare evolution. These results suggest that the initial impulsive phase developed under weakly collisional conditions. At later times, as the plasma density increased due to chromospheric evaporation, the simple relationships between the thermal and non-thermal electron populations disappeared, possibly indicating a transition to a different acceleration regime.

Let us consider the possible influence of chromospheric evaporation on electron acceleration by means of simple estimations. From the previous analysis of the MW and X-ray spectra at the peak of the non-thermal emission, we estimate that the fraction of accelerated electrons relative to the parent thermal plasma is approximately $n_b/n = 1.4\times 10^6/4.9\times 10^9\approx 3\times 10^{-4}$. We attempt to relate the temporal evolution of the accelerated-particle population to the chromospheric plasma flows and to the observed empirical relation $F(EM)$.

Consider a cylindrical magnetic loop whose diameter is much smaller than its length, with cross-sectional area (S) and length (L). Accelerated electrons, assumed to be energized in the corona, are injected toward the loop footpoints. Their precipitation produces plasma heating and an increase in pressure, which drives chromospheric evaporation and causes plasma to flow upward into the corona. Let ($n_{ch}$) be the chromospheric plasma density and ($v_{ch}$) the evaporation velocity. Conservation of particles then requires
$$ 2v_{ch}n_{ch}S = \frac{d}{dt}(nSL).$$
Since the partially ionized chromospheric plasma becomes fully ionized after entering the corona, the temporal evolution of the coronal density can be written as
$$\frac{dn}{dt} = 2\frac{v_{ch}}{L}n_{ch}\approx const $$
Considering the initial stage of the impulsive phase (the grey-shaded interval in Fig.~\ref{Xray_Energy}), we assume for simplicity that $n(t)\gg n_0$. In this case,
$$ \Delta n(t)=n(t)-n_0\approx n(t) \approx 2\frac{v_{ch}}{L}n_{ch}t $$
Thus, within these assumptions, the coronal plasma density increases approximately linearly with time, implying that $EM\propto F \propto t^2$. Using the observed relation $F(EM)$ together with the expression relating the accelerated-electron flux $F$ to the non-thermal electron density $n_b$, we obtain
\begin{equation}
\label{FE_con2} 
2n_bS\cdot\frac{\delta - 3/2}{\delta - 2}\cdot\sqrt{E_{low}\frac{2}{m_e}} \sim 2\gamma n^2SL.\nonumber
\end{equation}
A formal estimate of the mean velocity of accelerated electrons in the non-relativistic approximation is $$\langle V_{nth}\rangle \approx \frac{\delta - 3/2}{\delta - 2}\cdot\sqrt{E_{low}\frac{2}{m_e}} $$
For $E_{low}=45$~\,keV and $\delta = 5$, $\langle V_{nth}\rangle\approx c/2$, this yields
$$
n_b\langle v_{nth}\rangle
\sim
\gamma n^2L.
$$
Using the density $n(t)$ evolution determined by chromospheric evaporation, we obtain the following estimate for the fraction of accelerated electrons relative to the parent hot plasma:
$$
\frac{n_b}{n}
\sim
\gamma
\frac{j_{ch}}
{\langle v_{nth}\rangle}
t,
$$
where
$$
j_{ch}=n_{ch}v_{ch}
$$
is the chromospheric particle flux entering the loop.

Equivalently, the density of accelerated electrons can be expressed as
$$
n_b
\sim
\nu_{acc}
\frac{j_{ch}}
{\langle v_{nth}\rangle}
t.
$$

This result suggests that, during the early impulsive phase, the fraction of accelerated electrons may increase approximately linearly with time as chromospheric evaporation progressively supplies additional plasma to the coronal acceleration region.

Thus, it is shown that the fraction of accelerated electrons depends, first, on the properties of the accelerator itself, which are characterized by the parameters $\gamma$ and $<v_{nth}>$. The former parameter is related to the characteristic acceleration timescales and to the probability that a thermal particle will be accelerated. The latter depends on the minimum electron energy in the spectrum (the low-energy cutoff) and on the spectral index of the accelerated-electron distribution, both of which are determined by the physics of the acceleration process as well as by particle transport effects.

It was shown that, during the early impulsive phase, the transport regime along the entire loop is essentially collisionless, with $\lambda_{mfp}\gtrsim L/2$. Energy losses associated with the return current are also negligible, since
$E_{stop} = e\mathcal{E} = e f/\sigma\approx 0.04\,\mathrm{keV}$ for the peak of the X-ray emission \citep[]{ZharkovaGordovskyy2005}, where $\mathcal{E}$ is the return-current electric field and $\sigma$ is the plasma conductivity (we use thermal approach).

The absence of significant transport-related effects for most accelerated electrons is further supported by an estimate of the electron escape time from the loop toward the footpoints, $\tau_{esc}\sim n_bV/F\approx 0.02\,\mathrm{s}$~\citep[]{SharykinLiu2014}. This value is comparable, to within an order of magnitude, to the electron transit time across half of the loop, $\tau_{cross}\approx L/(2\langle v_{nth}\rangle)\approx 0.05\,\mathrm{s}.$ According to \citet{PetrosianChen2010}, $\tau_{esc}\approx \tau_{cross}[1+\tau_{cross}/\tau_{sc}]$, where $\tau_{sc}$ is the scattering time associated with Coulomb collisions and turbulent fluctuations. Since $\tau_{cross}\sim \tau_{esc}$ and $L\lesssim \lambda_{mfp}$, it follows that $\tau_{sc}\gg \tau_{cross}$, indicating a relatively low level of turbulence-induced scattering. We therefore suggest that the electron spectrum observed during the early impulsive phase represents a relatively “clean” acceleration spectrum, formed directly within the acceleration region and only weakly modified by transport effects in the magnetized plasma.

Secondly, the electron acceleration process appears to depend on the accumulation of plasma supplied to the corona through chromospheric evaporation with a particle flux $j_{ch}$. In other words, an increase in the plasma content of the magnetic loop leads to an increase in the number of accelerated electrons.

Using the relation derived above, we can estimate the evaporation velocity for (t=30) s (corresponding to the width of the grey-shaded interval in Fig.~\ref{Xray_Energy}):
$v_{ch} \approx n/(2n_{ch}\gamma t)\approx 7.5-75\,\mathrm{km~s^{-1}}$
assuming chromospheric electron densities in the range $10^{10}-10^{11}\,\mathrm{cm^{-3}}$. This velocity is comparable to or exceeds the chromospheric sound speed $\sim 7-15\,\mathrm{km~s^{-1}}$. The lower boundary of the estimated range appears more realistic, since the relatively hard electron spectrum $E_{low}\gtrsim 40$\,keV is expected to deposit most of its energy in the denser layers of the lower solar atmosphere. Therefore, during the early impulsive phase, we most likely observe a regime of the gentle chromospheric evaporation.

From this perspective, it is interesting to estimate what fraction of the energy carried by accelerated electrons is spent directly on driving the upward plasma flow into the corona. Let us consider an estimate of the amount of evaporated material $N_{evap}$ (expressed in units of $nL$, i.e. ($\mathrm{cm^{-2}}$) generated by chromospheric evaporation \citep[]{VeronigBrown2004,Fisher1989}, driven by accelerated electrons with an energy flux density $f(E>E_{low})$:
$$
N_{evap}\approx a_1\left[
a_2,
\beta
\left(
\frac{\delta}{2},
\frac{1}{3}
\right)
(\delta-2)
f(E>E_{low})
\right]^{\frac{2}{\delta-2}}
$$,
where $\beta$ denotes the beta function, and $a_1 = 8.2\times10^{19}$\,cm$^{-2}$ and $a_2=7.7\times10^{-12}$  are constant coefficients according to \citet{VeronigBrown2004}.

The condition derived above implies that chromospheric evaporation can develop only when the heating rate produced by the precipitating electron beam exceeds the local radiative loss rate around $T\approx 10^{5}$\,K, where the radiative loss function is approximately $f_R\approx7\times10^{-22}\,\mathrm{erg\cdot cm^{-3}\cdot s^{-1}}.$ Substituting the observed values, $f(E>E_{low})\approx8.5\times10^{9}\,\mathrm{erg\cdot cm^{-2}\cdot s^{-1}}$ and $\delta=5$, yields a critical column depth of $N_{evap}\approx3.5\times10^{19}\,\mathrm{cm^{-2}}$. The stopping column depth for electrons with energy $E_{low}$ is $N_{low}\approx2.61\times10^{20}\,\mathrm{cm^{-2}}$, so that $N_{low}>N_{evap}$.

In this case, the analytical treatment of \citet{Fisher1989} gives the fraction of the electron energy flux deposited into chromospheric evaporation:
$$\frac{f_{evap}}{f} = 1-\frac{\beta_x\left(\frac{\delta}{2}\text{, }\frac{1}{3} \right)}{3}x^{1-\frac{\delta}{2}} - (1-x)^{\frac{1}{3}}$$
where $x=N_{evap}/N_{low}$. For the present event we obtain $f_{evap}/f\approx0.04$.

Thus, only a small fraction $\approx4\%$ of the nonthermal electron energy is expected to be expended on driving chromospheric evaporation. This relatively low efficiency is mainly a consequence of the comparatively high low-energy cutoff of the accelerated-electron spectrum ($E_{low}\approx45$\,keV), which allows most electrons to penetrate into deeper atmospheric layers before depositing their energy.

Within the framework of these estimates, the energy partition during the early impulsive phase appears to be dominated by two comparable channels: direct plasma heating and electron acceleration. The energy required to drive chromospheric evaporation represents only a minor contribution to the overall energy budget. Since the flare was confined and no large-scale eruption was observed, the kinetic energy associated with bulk plasma motions is expected to be comparatively small and is therefore neglected in the present energy balance.

The estimates presented above suggest that, during the initial stage of the flare evolution (prior to the onset of the QPP phase discussed below), the particle acceleration process may have been modulated by chromospheric evaporation flows under nearly collisionless conditions within the reconnecting current sheet. In this scenario, an initial burst of accelerated electrons deposits energy into the lower atmosphere and drives an upward flow of chromospheric plasma. The arrival of this plasma at the reconnection region may perturb the current sheet and enhance the efficiency of subsequent electron acceleration.

The characteristic response time of the current sheet to the arrival of evaporated plasma can be estimated as $\tau_{resp}\approx \frac{L}{2c_s}\approx 8\,\mathrm{s}$, where $c_s\approx 990$\,km\,s$^{-1}$ is the sound speed in the flare loop (for $T=35.4$~MK). This timescale is comparable to the characteristic duration of the substructures observed during the early impulsive phase.

It is therefore possible that the three consecutive subpeaks observed in both the accelerated-electron flux and the emission measure within the shaded interval of Fig.\,\ref{Xray_Energy} represent episodes of impulsive chromospheric heating by nonthermal electron beams, followed by evaporation-driven feedback on the acceleration region. As the flare progresses, the pressure and density of the evaporated plasma within the loop increase substantially. Under such conditions, chromospheric evaporation is likely to become less effective in modulating the reconnection region, while collisional effects become increasingly important. Subsequently, the flare enters the phase characterized by pronounced QPPs, which dominate the later stage of the impulsive energy release.

\subsection{Possible Mechanisms of Quasi-Periodic Pulsations (QPPs)}

Before discussing possible mechanisms responsible for the observed QPPs, it is important to note that we cannot completely exclude the possibility that the analysed sequence of only 4--5 bursts, separated by similar but not identical time intervals, may have arisen by chance without any underlying quasi-periodic process. Nevertheless, statistically significant periodicities are detected independently by several methods, including Fourier analysis, wavelet analysis, and measurements of the time intervals between adjacent peaks. Moreover, these signatures are simultaneously present in observations from multiple instruments and spectral domains, including MW, HXR, and the time derivative of the SXR flux. Taken together, these results strongly suggest that the observed pulsations are associated with a physical mechanism rather than being a random sequence of impulsive bursts.

We also note that a sequence consisting of only 3--4 pulses may already be classified as a QPP event \citep{Nakariakov2019PPCF,Zimovets2021}. Similar short-duration pulse trains have been reported and discussed as manifestations of QPPs in numerous previous studies (e.g. \citealt{Nakariakov2010ApJ,LiDong2020ApJL,LiDong2024ApJ,Soloviev2025ARep}).

The observed QPPs exhibit several characteristic properties:
\begin{enumerate}
    \item [(a)] a small number of pulses (4--5 peaks) accompanied by rapid damping; 
    \item [(b)] non-harmonic pulse shapes, approximately triangular in form; 
    \item [(c)] signatures of non-stationarity, manifested by variations of the pulsation period around a mean value of $P_{QPP}=11.5\pm3.0$\,s, with a tendency to decrease from approximately 15\,s to 9\,s (by a factor of $\sim$1.7) during the short ($\sim$1\,min) impulsive phase of the flare; 
    \item [(d)] nearly simultaneous occurrence of peaks in HXR, MW, and the time derivative of the SXR flux, with no systematic delays exceeding the 1,s temporal resolution;
    \item [(e)] the presence of fine temporal structure (sub-peaks) within the major HXR and MW bursts; 
    \item [(f)] the appearance of QPP signatures only in non-thermal emissions (HXR, MW, and the SXR derivative), while no clear periodic component is detected in the thermal SXR emission below 10--12\,keV; 
    \item [(g)] systematic shifts of the apparent MW source brightness centroid from pulse to pulse, corresponding to projected velocities in the range $v_{MW}\approx200$--800\,km\,s$^{-1}$.  
\end{enumerate}
Additional observational evidence includes the extended and spatially non-uniform structure of the UV flare ribbons located on opposite sides of a strongly curved PIL, as well as the complex morphology of EUV loops and the extrapolated coronal magnetic field, both of which indicate a multi-loop magnetic configuration of the flare region. Taken together, these observations suggest that the detected QPPs are most likely associated with a sequence of magnetic reconnection episodes and successive injections of accelerated electrons occurring within a complex magnetic environment (see reviews of QPP mechanisms and their observational signatures in \citealt{NakariakovMelnikov2009,Kupriyanova2020,Zimovets2021,Reale2026}). The principal unresolved question concerns the physical mechanism responsible for the quasi-periodicity itself and for setting the characteristic timescale of the pulsations.

Mechanisms based on the modulation of non-thermal electromagnetic emission by MHD oscillations of a single flare loop, as well as oscillations of the loop treated as an LRC circuit, are considered unlikely in the present event because of the complex multi-loop structure of the flare UV/EUV sources and the reconstructed coronal magnetic field.  We therefore examine whether standing fast or slow magnetoacoustic modes of flare loops could modulate the energy release and electron acceleration processes through quasi-periodic modulation of the reconnection efficiency in the vicinity of a magnetic null point or current sheet \citep{Nakariakov2006AA}.

The period of a standing mode is determined by the ratio of twice the loop length $L$ to the corresponding phase speed $C_{ph}$, i.e., $P=2L/C_{ph}$. The lengths of the flare loops in the studied region are confined to a relatively narrow range, $L=(1.5$--$3.2)\times10^{9}$\,cm. The phase speed of the slow magnetoacoustic mode can be approximated by the sound speed in the loop, $C_{sma}\approx C_s \approx 1.66\times10^4 T^{1/2} \approx 9.9\times10^7$\,cm\,s$^{-1}$ (or $\approx990$\,km\,s$^{-1}$), for the plasma temperature $T=35.4$\,MK inferred from the X-ray spectral analysis.

The phase speed of the fast magnetoacoustic mode may be estimated as $ C_{fma}\approx \sqrt{C_s^2+C_A^2}$, where the Alfvén speed is $C_A\approx 2.18\times10^{11}Bn^{-1/2} \approx (1.7\text{--}2.2)\times10^9$\,cm\,s$^{-1}$ (or $\approx(1.7\text{-}2.2)\times10^{4}$\,km\,s$^{-1}$) for magnetic field strengths $B=547$--693\,G and a plasma density of $n=4.9\times10^9$\,cm$^{-3}$. Consequently, the estimated phase speed of the fast magnetoacoustic mode is $C_{fma}\approx(1.7$--$2.2)\times10^9$\,cm\,s$^{-1}$, nearly twenty times greater than $C_{sma}$. These values are substantially higher than those typically reported for flare-surrounding coronal loops \citep{Nechaeva2019ApJS,Nakariakov2021SSR}, owing to the strong magnetic field and relatively low plasma density of the flare loops considered here.

The corresponding periods of the fundamental slow and fast magnetoacoustic modes (e.g. standing kink or trapped sausage modes) are estimated to be $P_{sma}\approx30.3\text{--}64.6\,\mathrm{s}$ and $ P_{fma}\approx1.4\text{--}3.8\,\mathrm{s} $,
respectively. The former exceeds the observed QPP period by a factor of approximately 2--4, whereas the latter is smaller by a factor of about 4--11. Thus, both estimates differ significantly from the observed value.

For the leaky sausage mode, the fundamental period is controlled primarily by the loop radius rather than by its length. Since the loop radius is at least several times smaller than the loop length (see Section~5.1), the corresponding period would be substantially shorter than the estimate obtained above for the trapped sausage mode.

These order-of-magnitude estimates therefore argue against standing loop oscillations as the primary mechanism modulating the efficiency of electron acceleration and/or injection in this event. Additional observational arguments against this class of models are provided by properties (b)--(g) listed above, particularly the pronounced non-stationarity of the pulsations, the systematic displacement of the MW source centroid, and the complex multi-loop magnetic structure of the flare region (see also \citealt{Zimovets2021}).

Another possible explanation for the observed QPPs in the non-thermal electromagnetic emission is oscillatory magnetic reconnection \citep{McLaughlin2018SSR, Karampelas2023ApJ, Stewart2022MNRASb}. Oscillatory reconnection is a physical process characterized by a series of reconnection episodes accompanied by periodic changes in the magnetic connectivity of the disturbed magnetic field. An important property of oscillatory reconnection is that the periodicity is not imposed by an external driver, as in the case of loop oscillations discussed above, but is instead an intrinsic property of the relaxation process itself. Furthermore, the oscillation period is predicted to be largely independent of the amplitude of the initial triggering disturbance \citep{Karampelas2022ApJ}.

Numerical simulations of oscillatory reconnection frequently produce electric-current density profiles near magnetic null points consisting of a small number (typically 4--7) of damped peaks. Such behaviour may lead to the development of plasma instabilities and anomalous resistivity and, consequently, to quasi-periodic variations in the reconnection rate and accelerating electric field. In turn, this could result in a short series of decaying QPPs associated with repeated episodes of electron acceleration. It should be noted, however, that neither electron acceleration nor the resulting electromagnetic emission was modelled explicitly in these studies, to the best of our knowledge. Therefore, the following discussion remains necessarily speculative.

Based on a parametric study of a series of numerical MHD simulations, \citet{Karampelas2023ApJ} derived an approximate expression (their Equation~17) relating the oscillation period to the background magnetic field strength, plasma temperature, and density in the vicinity of a magnetic null point. Substituting the physical parameters inferred in the present study, namely $B=650$\,G, $n=4.9\times10^{9}$\,cm$^{-3}$, and $T=35.4$\,MK, yields an oscillation period of approximately $P \approx 47.5 \pm 3.4\,\mathrm{s} $, which exceeds the observed QPP period by a factor of about 3--4.

Several caveats should be noted, however. The simulations of \citet{Karampelas2023ApJ} were performed for an idealized two-dimensional X-point configuration without a guide field and for a limited range of plasma parameters: magnetic field strengths of 10--30\,G, plasma temperatures of 5--10\,MK, and densities of approximately $(1$--$20)\times10^{9}$\,cm$^{-3}$. Both the magnetic field strength and temperature in these models are substantially lower than the values inferred for the present flare. Moreover, our estimates correspond to flare loops after plasma heating and electron acceleration have already occurred, whereas the model parameters describe pre-reconnection coronal conditions near the X-point.

Using the same density, $n=4.9\times10^{9}$\,cm$^{-3}$, but reducing the magnetic field strength to 30\,G and the temperature to 5\,MK yields a predicted oscillation period of $ P \approx 50.8 \pm 3.4\,\mathrm{s} $, indicating that the model period changes only weakly over this parameter range. However, periods in the range of approximately 13--16\,s, comparable to the observed QPP period, can be obtained for parameter combinations such as $B=10$--30\,G, $T=5$--10\,MK, and $n=1.2\times10^{9}$\,cm$^{-3}$. In this case, the plasma density near the X-point is approximately four times lower than that within the flare loops, which have been substantially filled by chromospheric evaporation. Such a difference appears entirely plausible given the uncertainties of the present estimates. Likewise, the magnetic field strength near a coronal null point may reasonably be expected to be several times lower than that within the flare loops into which accelerated particles are injected \citep{Chen2020NatAs, Edgar2024MNRAS}.

Therefore, oscillatory reconnection cannot be ruled out as a possible mechanism operating in the present event. Nevertheless, it remains unclear whether this mechanism can account for the observed decrease of the QPP period with time and the systematic displacement of the MW source centroid. For example, MHD simulations of oscillatory reconnection triggered by emerging magnetic flux \citep{Murray2009AA, WangYifu2025ApJ} predict a considerably more complex evolution of flare sources than observed in the present event. However, such complexity may simply remain unresolved because of the limited temporal and spatial resolution of the available observations.

Systematic displacements of flare emission sources observed in SXR, HXR, MW, EUV/UV, and optical wavelengths have been reported in numerous studies, including during QPP episodes (e.g., \citealt{Grigis2005ApJ, Zimovets2009SoPh, Kim2013PASJ, LiZhang2015ApJ, Reva2015SoPh, Kuznetsov2016SoPh, Kuznetsov2017GiA, Purkhart2025AA, Lorincik2026ApJ, LiTing2026arXiv}). In most cases, the apparent source velocities range from several tens to several hundreds of kilometres per second, comparable to typical slow magnetoacoustic wave speeds under coronal plasma conditions. Such apparent motions are frequently interpreted in terms of slipping magnetic reconnection occurring within quasi-separatrix layers characterized by strong gradients in magnetic connectivity. However, a convincing explanation for the quasi-periodicity of such reconnection remains lacking.

Several mechanisms have been proposed to account for the observed source displacements. These include modulation of reconnection by slow magnetoacoustic waves repeatedly reflected from the chromosphere \citep{NakariakovZimovets2011ApJ}, modulation by flapping waves in the current sheet \citep{ArtemyevZimovets2012SoPh}, and thermal-instability waves developing within the reconnecting current sheet \citep{Ledentsov2021aSoPh, Ledentsov2021bSoPh}.

For the flare studied here, the apparent displacement velocity of the MW source was estimated to be approximately $200$--$800$\,km\,s$^{-1}$. This range is comparable to the estimated slow magnetoacoustic speed, $ C_{sma}\approx 990\,\mathrm{km\,s^{-1}} $, for a plasma temperature of $T=35.4$\,MK, and remains an order of magnitude smaller than the estimated fast magnetoacoustic speed (see discussion above). For this reason, the scenario in which slow magnetoacoustic waves periodically trigger reconnection \citep{NakariakovZimovets2011ApJ} appears more plausible than the alternative mechanisms considered here.

Within this framework, the observed decrease of the QPP period may be interpreted as a consequence of a gradual reduction in the height of the reconnecting current sheet (or, equivalently, the characteristic loop length) and/or an increase in the slow magnetoacoustic speed. The latter could naturally occur as the flare plasma temperature rises during the course of the event. Nevertheless, the compactness of the flare region, together with the limited spatial and temporal resolution of the available observations, does not allow a more rigorous test of this scenario \citep[see also][]{InglisDennis2012ApJ, Zimovets2021SoPh}.

We also consider the mechanism proposed by \citet{Emslie1981ApL}, in which the formation of current sheets and the triggering of successive reconnection episodes are driven by the lateral expansion of flare loops filled with heated plasma and their interaction with neighbouring magnetic structures. This mechanism belongs to the class of auto-wave processes and could, in principle, explain a quasi-periodic sequence of electron injections occurring in adjacent loops.

However, for this scenario to operate, the flare loop must first become unstable and expand sufficiently to interact laterally with neighbouring loops. A necessary condition for such an expansion is that the plasma beta, $ \beta=p_{th}/p_{m}$, defined as the ratio of thermal plasma pressure to magnetic pressure, reaches values of the order of unity or higher, i.e. $ \beta \gtrsim 1 $. Substituting the physical parameters inferred for the present event into the standard expression for the plasma beta yields $ \beta \approx 6.94\times10^{-15} nTB^{-2}
\approx (2.5-4)\times10^{-3} $, which is more than two orders of magnitude smaller than unity. Therefore, the flare loops are expected to remain strongly magnetically confined, making substantial lateral expansion unlikely.

An additional difficulty arises when comparing the predicted and observed periods. In the model of \citet{Emslie1981ApL}, the characteristic QPP period is given by $ P_{\rm Emslie}=D/C_{A}$, where $D$ is the characteristic separation between interacting loops and $C_A$ is the Alfvén speed. In the flare studied here, the distance between neighbouring flaring loops does not exceed the loop length itself, i.e. $ D<L $ (see Fig.~\ref{NLFFF}). Given the high Alfvén speed inferred for the flare loops (see discussion above), this relation inevitably implies $ P_{\rm Emslie}<P_{\rm QPP} $, in disagreement with the observed QPP period.

The model also faces a more fundamental difficulty. It remains unclear what physical mechanism would establish a characteristic spacing between successively activated flare loops. Magnetic field and plasma are distributed continuously, albeit non-uniformly, throughout the AR, and therefore the existence of a preferred and approximately constant distance between neighbouring reconnection sites is not obvious. Taken together, these considerations argue against the applicability of the \citet{Emslie1981ApL} mechanism to the present event.

We consider QPP mechanisms associated with processes occurring within twisted magnetic flux ropes~\cite[e.g.,][]{Zimovets2018JASTP, Smith2022MNRAS, Stewart2025MNRAS, Soloviev2025ARep} to be unlikely in the present event, as the NLFFF extrapolation does not reveal any clear flux-rope structures in the form of coherent bundles of magnetic field lines twisted about a common axis. Instead, the reconstructed magnetic field is characterised by a complex system of low-lying loops exhibiting varying degrees of shear and different orientations with respect to the PIL.

Finally, based on the available observations, we must also consider another possible mechanism associated with the nonlinear coalescence instability of current-carrying loops \citep{Tajima1987}. This mechanism has been invoked in several studies to interpret nonthermal QPPs in solar flares with periods of approximately $1$--$10$\,s, including a number of C-class events (see references in the Introduction).

A simple estimate of the characteristic oscillation period can be obtained by considering the Alfvén transit time across the magnetic flux tube cross-section, $ P \sim \tau_{A} = a/C_{A} $. Substituting the parameters inferred for the present flare, $ a \approx 3.5\times10^{8}$\,cm and $ C_{A} \approx (1.7-2.2)\times10^{9}$\,cm\,s$^{-1}$, yields
$ P \approx 0.2$\,s, which is nearly two orders of magnitude shorter than the observed QPP period.

An even more stringent constraint follows from the expression for the minimum oscillation period of the current sheet during loop coalescence derived by \citet{Tajima1987}, $ P \approx 2\pi \beta^{3/2}\tau_A $. Using the estimated plasma beta values obtained above gives $ P \approx 4\times10^{-3}\,\mathrm{s} $, which is substantially smaller than the observed periodicity.

These unusually short model periods arise primarily from the very high Alfvén speed in the flaring loops. In turn, the large Alfvén speed is a direct consequence of the strong magnetic field inferred from the NLFFF reconstruction, $ B \approx 550-700\,\mathrm{G} $, combined with the relatively low plasma density. As a result, the nonlinear coalescence instability predicts characteristic timescales that are far shorter than those observed. Therefore, this mechanism appears unlikely to be responsible for the QPPs detected in the present event.

In summary, despite the rich observational dataset available for this event, we are unable to draw a definitive conclusion regarding the physical mechanism responsible for the observed QPPs. Nevertheless, the collective observational evidence strongly suggests that the QPPs represent a short series of episodic magnetic reconnection events accompanied by repeated injections of accelerated electrons into the complex system of flare loops.

The key unresolved question concerns the trigger responsible for the quasi-periodic modulation of the reconnection process. Among the mechanisms considered in this study, the most plausible candidates are oscillatory reconnection and periodic modulation of reconnection by slow magnetoacoustic waves reflected from the chromosphere and propagating along the system (i.e. a highly sheared arcade) of the flare loops. Both mechanisms are capable, at least qualitatively, of producing a short sequence of nonthermal pulsations with characteristic periods comparable to those observed.

However, the spatial and temporal resolution of the available observations, together with the compact and morphologically complex structure of the flare region, do not allow a more rigorous discrimination between these scenarios. Further progress will likely require observations with substantially improved temporal and spatial resolution, as well as dedicated numerical modelling tailored to the magnetic configuration of the event.

\subsection{Polarization Properties and Spectral Characteristics in the Radio Domain}

An important observational feature of the studied event is the reversal of the circular polarization sign (Stokes~$V$ parameter) in the MW range. The combined analysis of SOLARSPEL, NoRP, and SRH observations confirms the reality of this effect: a change in the polarization sign is observed near frequencies of 6--7\,GHz. At frequencies below 6\,GHz, the Stokes~$V$ component is positive, whereas at higher frequencies the observed polarization becomes negative.

Within the framework of the classical theory of gyrosynchrotron emission, such a polarization reversal may be associated either with the transition from the optically thick to the optically thin emission regime or with propagation effects in an inhomogeneous magnetic field, particularly quasi-transverse mode coupling. In the first case, the sign reversal occurs naturally near the spectral peak, where the dominant magnetoionic mode changes as the optical depth decreases. In the second case, propagation through quasi-transverse regions of the coronal magnetic field can alter the observed polarization mode and lead to a reversal of the Stokes~$V$ sign.

In the present work, we limit ourselves to establishing the existence of the polarization inversion on the basis of observational evidence. A detailed investigation of the physical mechanisms responsible for this phenomenon would require dedicated modelling of both the gyrosynchrotron source and the propagation of radio waves through the surrounding coronal plasma and magnetic field. Such an analysis is beyond the scope of the present study and is left for future work. Possible interpretations of polarization reversals in solar MW bursts are discussed in detail by \citet{Zheleznyakov1970} and \citet{White1992}.

Another noteworthy characteristic of the event is the absence of a significant radio response in the metric and decimetric wavelength ranges. Analysis of observations from the e-Callisto network and the Learmonth radio spectrograph revealed no detectable low-frequency radio bursts associated with the studied flare. Data from the YAMAGAWA spectropolarimeter also remained at the noise level and therefore could not be used for quantitative analysis.

The lack of radio emission at metric wavelengths indirectly supports the compact and confined nature of the event. In particular, it suggests the absence of large-scale eruptive processes and indicates that accelerated electrons were not efficiently injected into extended coronal structures or escaping magnetic field lines. This interpretation is consistent with the magnetic field topology inferred from the NLFFF extrapolation and with the absence of observational signatures of a coronal mass ejection or other large-scale eruptive phenomena.

\subsection{Nature of the Short-Duration Narrowband Burst}

A narrowband radio burst was detected near the peak of the MW emission in the frequency range 4.0--4.6\,GHz (Fig.~\ref{MWspect}). Several of its properties suggest a coherent emission mechanism, including its narrow spectral bandwidth, substantially higher intensity and degree of polarization compared to the main MW component, and its complex, highly structured spectral morphology.

The characteristic delay of approximately 4\,s between this burst and the main MW peak may reflect the development time of a plasma instability within the source region.

Assuming a plasma emission mechanism operating near 5\,GHz, the corresponding electron density can be estimated as $ 
n_e \approx 2.5\times10^{11}\ {\rm cm^{-3}} $. If the emission occurs at the second harmonic of the plasma frequency, the inferred density is four times lower, $ n_e \approx 0.6\times10^{11}\ {\rm cm^{-3}} $. Both estimates substantially exceed the plasma density derived from the X-ray source analysis $\approx 0.5\times10^{10}\,\mathrm{cm^{-3}}$. This discrepancy may indicate that the radio emission was generated in significantly denser layers of the solar atmosphere, for example near the footpoints of magnetic loops or in a current sheet associated with very low-lying compact loops, which we suggest were involved in the initial stage of the flare.

Additional support for a dense source region follows from the conditions required for the excitation of Langmuir waves within the framework of quasilinear theory \citep[e.g.,][]{KaplanTsytovich1973, Kontar2001}. The generation of Langmuir turbulence becomes possible when the growth rate of the beam instability exceeds the rate of thermal Coulomb collisions \citep{KaplanTsytovich1973}, $ \frac{n_b}{n}\left(\frac{v_b}{\Delta v_b}\right)^2 \sim \frac{1}{N_d} $, where $N_d$ is the number of electrons within a Debye sphere and $v_b/\Delta v_b$ characterizes the velocity dispersion of the electron beam.

Using order-of-magnitude estimates, $ \left(\frac{v_b}{\Delta v_b}\right)^2 \sim 10 $ and $ \frac{n_b}{n} \sim 10^{-4} $,
one obtains $ N_d^{-1}\sim10^{-3} $. In contrast, for the flaring coronal plasma with a temperature of approximately 40\,MK and a density of $4\times10^{9}\,\mathrm{cm^{-3}}$, the corresponding value is only $ N_d^{-1}\sim10^{-10} $. In denser and cooler chromospheric plasma, where both $N_d$ and $n_b/n$ are significantly different, the instability criterion may be satisfied more readily. However, these considerations are meaningful only insofar as the quasilinear approximation remains applicable.

A comparison between SOLARSPEL and SRH observations reveals notable differences in the temporal evolution of the Stokes~\textit{V} parameter. Whereas the SRH data in the 6--12\,GHz range exhibit a single polarization reversal, the SOLARSPEL observations show a more complex behaviour with multiple sign changes of the circular polarization component. This difference is most likely related to the higher temporal and spectral resolution of SOLARSPEL, which allows the detection of fine structures that are partially averaged out in the SRH data due to its scanning cadence of approximately 3\,s.

An alternative explanation for the observed narrowband highly polarized emission is the electron cyclotron maser (ECM) mechanism \citep{Melrose1982}. Efficient ECM generation, however, requires rather restrictive plasma conditions, in particular $ \omega_{pe}/\Omega_{ce}<1 $, which may be difficult to satisfy in the presence of surrounding dense plasma capable of absorbing or suppressing the escaping radiation.

In summary, the detection of a coherent radio burst produced by accelerated electron beams in dense atmospheric layers during a short interval of the impulsive phase is of considerable interest. One possible interpretation is that two spatially distinct acceleration regions were present during the flare: (1) a coronal source responsible for the gyrosynchrotron MW emission and (2) a lower-altitude source producing the coherent narrowband burst. The relationship between these potential acceleration sites remains unclear. Equally plausible is a scenario involving a single acceleration region undergoing a transient change in its acceleration regime or emission conditions.

A more comprehensive investigation of the origin of this narrowband burst, both in the present event and in similar flares, will require observations with substantially improved temporal, spectral, and spatial resolution, together with dedicated theoretical modelling.

\section{Conclusions}
\label{sec:conclusions}
In this work, we presented a detailed multiwavelength analysis of the non-eruptive impulsive C2.8 solar flare that occurred on 19 March 2023. Despite its relatively modest GOES class, the event exhibited a complex morphology and energy-release pattern comparable to those observed in larger flares. The absence of an eruption and large-scale plasma ejections simplifies the interpretation of the observations and provides an opportunity to investigate the early stages of flare energy release under relatively confined conditions.

The combined analysis of X-ray and MW observations enabled quantitative diagnostics of accelerated electrons and their relationship to the properties of the heated plasma during the rise phase of the impulsive stage. We identified empirical relationships between the total flux of nonthermal electrons and the plasma emission measure, as well as between the plasma temperature and the low-energy cutoff of the accelerated-electron spectrum. We also found approximate energy equipartition between the total energy of accelerated electrons and the thermal energy of the hot plasma. These results provide a rare opportunity to investigate particle acceleration directly from a thermal plasma population before chromospheric evaporation and other flare hydrodynamic processes become fully developed.

Our estimates suggest that the efficiency of electron acceleration is relatively low. Under the conditions inferred for the early impulsive phase, a thermal electron requires, on average, approximately $10^2$\,s to become part of the nonthermal power-law population above $\sim40$\,keV. The observed relationships between thermal and nonthermal parameters further indicate that the acceleration process may be directly modulated by the supply of plasma through chromospheric evaporation. A simple estimations were developed linking the fraction of accelerated particles to both the accelerator efficiency and the chromospheric plasma inflow.

The reconstructed magnetic-field geometry reveals a complex system of low-lying magnetic loops rooted near the flare UV ribbons. The absence of any eruption suggests that energy release occurred within confined current-sheet structures characterized by a substantial guide-field component rather than in a classical two-dimensional reconnecting configuration. The event therefore provides valuable quantitative constraints on the earliest stages of magnetic energy release, plasma heating, and particle acceleration under highly confined conditions. The relatively low acceleration efficiency inferred from the observations may be related to the properties of these guide-field-dominated current sheets.

The impulsive phase was accompanied by nonstationary QPPs observed in nonthermal emissions (MW, hard X-rays, and the time derivative of soft X-rays) with an average period of $11.5 \pm 3.0$\,s. The pulsations exhibited both amplitude damping and a gradual decrease in period from approximately 15 to 9\,s. Although the available observations do not allow a unique identification of the underlying mechanism, the evidence suggests that the QPPs represent a short sequence of 4--5 episodic reconnection events accompanied by repeated injections of accelerated electrons into a complex multi-loop magnetic system. Among the mechanisms considered, oscillatory magnetic reconnection and quasi-periodic modulation of reconnection by chromospherically reflected slow magnetoacoustic waves appear to be the most plausible candidates.

Finally, this compact confined flare demonstrates both the capabilities and the current limitations of modern solar observing facilities, including SRH, SOLARSPEL, SDO, GOES/XRS, Fermi/GBM, SolO/STIX, and ASO-S/HXI. The combination of observations across multiple wavelength ranges provides powerful constraints on flare energy release and particle acceleration processes. However, the available spatial and temporal resolution remains insufficient to fully resolve several key questions, including the trigger mechanisms of flare energy release, the origin of QPPs, the details of particle acceleration, and the partitioning of released magnetic energy. Addressing these problems will require both next-generation observations and the development of advanced data-driven flare models that combine realistic magnetic-field geometries, MHD simulations, particle transport, and synthetic emission calculations across the electromagnetic spectrum.

\section*{ACKNOWLEDGMENTS}
\label{sec:acknowledgments}
The authors gratefully acknowledge the teams of GOES, SDO, Fermi/GBM, Solar Orbiter/STIX, ASO-S/HXI, YAMAGAWA, NoRP, and e-Callisto for providing the open-access observational data and software used in this study. We are particularly grateful to the Radio Astrophysics Department of the Institute of Solar-Terrestrial Physics SB RAS for providing the SRH and SOLARSPEL observations, as well as for their scientific and technical support. I.V.Z. thanks Prof. W.-Q. Gan for his hospitality in PMO CAS and the opportunity to work with ASO-S data as part of the ASO-S Guest Investigator Program (AGIP).

The authors also acknowledge Research and Production Company Micran Joint Stock Company (Tomsk, Russia) for its contribution to the development and construction of the SRH and the SOLARSPEL.

This work was carried out using data obtained with the Unique Scientific Facility ``Radioheliograph'' operated by the Institute of Solar-Terrestrial Physics SB RAS\footnote{\url{https://ckp-rf.ru/catalog/usu/4138190/}}.
\section*{FUNDING}
This work was supported by the Russian Science Foundation (Grant No. 25-22-00745).

\section*{CONFLICT OF INTEREST}
The authors declare that they have no conflict of interest.

\bibliographystyle{aspb1}

\end{document}